\documentclass[a4paper,onecolumn,11pt, titlepage, accepted=2020-04-19]{quantumarticle}
\pdfoutput=1
\usepackage[utf8]{inputenc}
\usepackage[english]{babel}
\usepackage[T1]{fontenc}
\usepackage{amsmath}
\usepackage{amssymb}
\usepackage{amsthm}
\usepackage{hyperref}
\usepackage{physics}
\usepackage{float}
\usepackage{graphicx}
\usepackage{ulem}
\usepackage{multirow}
\graphicspath{ {figures/} }

\usepackage{tikz}
\usepackage{lipsum}

\usepackage{color}

\newtheorem{theorem}{Theorem}
\newtheorem{conjecture}[theorem]{Conjecture}
\newtheorem{proposition}[theorem]{Proposition}

\newtheorem{remark}[theorem]{Remark}

\begin{document}
	
	\title{Bipartite quantum measurements 
		with optimal single-sided distinguishability}
	
	\author[1]{Jakub Czartowski}
	\email{jakub.czartowski@doctoral.uj.edu.pl}
	\orcid{0000-0003-4062-833X}
	\author[1,2,3]{Karol Życzkowski}
	\orcid{0000-0002-0653-3639}
	\affil[1]{Faculty of Physics, Astronomy and Applied Computer Science, Jagiellonian University, ul. Łojasiewicza 11, 30-348 Kraków, Poland}
	\affil[2]{Centrum Fizyki Teoretycznej PAN, Al. Lotników 32/46, 02-668 Warszawa, Poland}
	\affil[3]{National Quantum Information Center (KCIK), University of Gda{\'n}sk,
		Poland}
	\maketitle
	\begin{abstract} 
		We analyse orthogonal bases in a composite $N\times N$ Hilbert space describing a bipartite quantum system and look for a basis with optimal single-sided mutual state distinguishability. This condition implies that in each subsystem the $N^2$ reduced states form a regular simplex of a maximal edge length, defined with respect to the trace distance. In the case $N=2$ of a two-qubit system our solution coincides with the {\sl elegant joint measurement} introduced by 
		Gisin. We derive explicit expressions of an analogous constellation for $N=3$ and provide a general construction of $N^2$ states forming such an optimal
		basis in ${\cal H}_N \otimes {\cal H}_N$. 	Our construction is valid for all dimensions
		for which a symmetric informationally complete (SIC) generalized measurement is known.   
		Furthermore, we show that the one-party measurement that distinguishes the states of an optimal basis of the composite system leads to a local quantum state tomography with a linear reconstruction formula. Finally, we test the introduced tomographical scheme on a complete set of three mutually unbiased bases for a single qubit using two different IBM machines.
	\end{abstract}

	\section{Introduction}
	
	According to Asher Peres {\sl quantum phenomena do not occur in a Hilbert space, they occur in a laboratory} \cite{Pe95}. However, the vectors used as a tool to compute probabilities describing outcomes of a quantum measurement do live in a complex Hilbert space.
	
	The Hilbert space ${\cal H}_N$ is isotropic, hence `all quantum states are equal'. On the other hand, if we consider $k=2$ pure quantum states, $\ket{\phi}$ and $\ket{\psi}$ of a fixed finite dimension $N$, the situation changes, as such a particular pair of states can be characterized by their fidelity, $F=\abs{\braket{\psi}{\phi}}^2$. In particular, if $F=0$ both states are orthogonal, and they are perfectly distinguishable. Therefore some {\sl configurations} of $k$ quantum states become `more equal than the others'.
	
	For a larger number $k$ of pure states in a given constellation the space of all their possible configurations grows fast with the number $k$ and the dimension $N$. Current interest in discrete structures in the Hilbert space \cite{BZ17} is driven by their importance for the foundations of quantum theory, but also by numerous applications in quantum information processing. On the one hand, we wish to improve our understanding of the structure of the set of quantum states and establish existence of certain distinguished constellations of states for a given dimension. On the other hand, one would like to find optimal schemes of quantum measurements which ensure the best way to extract information from the experimental data \cite{Pe95}.
	
	In particular, designing a scheme of an optimal quantum measurement plays a crucial role in various setups motivated by tests of Bell inequalities \cite{B66, GB98}, quantum cryptography \cite{BB84, GRTZ02}, quantum tomography and state identification \cite{MP95, S06} and studies of quantum networks, which display non-multilocal properties \cite{G17, G19, TGB20}. The simplest measurement scheme of L{\"u}ders and von Neumann can be associated to a given non-degenerate observable, which defines a basis consisting of $N$ orthogonal states.
	
	A more advanced scheme is given by a set of $N^2$ pure states with constant overlaps, which form a simplex of dimension $N^2-1$ inscribed into the set of all mixed states of dimension $N$. Such a generalized measurement is symmetric and {\sl informationally complete} \cite{RBKSC04}, as it allows one to extract all $N^2-1$ parameters describing a mixed state. This scheme, often called SIC in the literature, is proven to be optimal for quantum state tomography \cite{S06}. Such a particular generalized measurements known for a fast-growing number of  dimensions \cite{S17,ACFW18,GS17, AB19, Gun}, but its existence for an arbitrary $N$ has not yet been established \cite{FHS17,B20}. 
	
	A link between orthogonal measurements and SICs was first demonstrated by Massar and Popescu \cite{MP95}, who proposed a game concerning guessing the expected direction of a set of several parallel spins. An analogous problem concerning quantum information for antiparallel spins was analysed in \cite{GP99}. These results allowed Gisin to identify a bipartite basis with tetrahedral symmetry called the {\sl Elegant Joint Measurement} (EJM) \cite{G17}. This construction was later studied in the context of quantum networks \cite{G19} and extended to an entire family of configurations with different degrees of entanglement \cite{TBGR20}.
	
	The aim of this work is to analyse orthogonal bases in a composite Hilbert space ${\cal H}_N \otimes {\cal H}_N$ describing a bipartite quantum system identify a particular basis such that it provides the maximal local mutual single-sided distinguishability. We assume here that the measurements are performed locally in both subsystems without any exchange of information, so that both parties have only an access to the information carried by reduced density matrices. According to the Helstrom theorem \cite{H69}, in order to identify a configuration ensuring the largest single-sided distinguishability one needs to find a set of $N^2$ bipartite orthogonal pure states such that in both subsystems the trace distance between any two partial traces is equal and becomes as large as possible.
	
	The key feature of the discussed setup is the required symmetry between both subsystems -- the distinguishability concerns reduced states analysed separately by both parties. If one considers the reduced density matrices of a single party only, one arrives at an implementation of a rescaled SIC by an orthogonal measurement in a larger space \cite{DJR05}.
	
	The paper is organized as follows. In Section \ref{sec:scene} we review the necessary notions and recall the Helstrom theorem. In Section \ref{sec:bases} the main result of the work is presented: an explicit construction of the desired basis with the optimal single-sided distinguishability. In the case $N=2$ a connection to the elegant joint measurement \cite{MP95, G19} is demonstrated. In Section \ref{sec:EJM3} we show explicitly a direct analogue of this measurement for two qutrits. Possible applications of the analysed measurement schemes to the quantum state tomography are discussed in Section \ref{sec:experiment}, in which their practical implementations on IBM quantum computers are presented. Finally, in Section \ref{sec:conclude} we conclude the work and present some related open problems.
	
	\section{Setting the scene} \label{sec:scene}
	
	\subsection{SIC measurements and complex projective designs}
	
	Symmetric informationally complete (SIC) generalized measurement belongs to the key notions analysed in this work. A SIC is defined as a set $\qty{\ket{i}}_{i=1}^{N^2}$ of $N^2$ pure states in ${\cal H}_N$, with a constant overlap \cite{RBKSC04},
	
	\begin{equation}
		\abs{\ip{i}{j}}^2 = \frac{1 + N\delta_{ij}}{1+N}.
	\end{equation}
	In the simplest case of $N=2$ such a construction is given by four quantum states corresponding to the vertices of a regular tetrahedron inscribed in the Bloch sphere. Zauner conjectured \cite{Z99, Z11} that such ensembles exist for any $N$. Analytic constructions are known \cite{Gun, ACFW18} for all dimensions $N \leq 53$, while numerical solutions were obtained \cite{S17, FHS17, S19} for dimensions $N \leq 193$, with some additional solutions known for certain dimensions \cite{GS17, AB19, B20} up to $N = 2208$.
	
	To introduce a more general notion let us consider an ensemble of $m$ pure states $\qty{\ket{\phi_i}}_{i = 1}^m$ and fix a natural number $t$. Such a set is called a {\sl complex projective $t$-design}, if the integral of any polynomial function $f_t(\ket{\phi})$ of degree $t$ in the components of the state $\ket{\phi}$ and their conjugates over the entire space of the pure states $\mathcal{H}$ with respect to the uniform, Fubini-Study measure $\dd{\ket{\phi}_{\text{FS}}}$ is equal to the average with respect to the ensemble, 
	\begin{equation}
		\frac{1}{m}\sum_{i=1}^m f_t(\ket{\phi_i}) = \int_\mathcal{H}\dd{\ket{\phi}}_\text{FS} f_t(\ket{\phi}) .
	\end{equation}
	Scott has proven that measurement schemes based on 2-designs are optimal for the task of quantum state tomography, providing a linear formula for reconstruction and minimizing the errors introduced by state preparation and  imperfections of the measurement \cite{S06}. Interestingly, any SIC measurement provides a minimal example of a complex projective 2-design, which consist of $N^2$ states of dimension $N$.

	A similar structure may also be defined for mixed quantum states \cite{CGGZ20}. An ensemble of $m$ mixed states $\qty{\rho_i}_{i=1}^m$ is called a \textbf{mixed-state $t$-design}, if the average of $t$ copies of the states in the ensemble is equal to the average over the entire space $\Omega_N$ of mixed quantum states of size $N$ with respect to the flat Hilbert-Schmidt measure $\dd{\rho}_\text{HS}$,
	\begin{equation}
		\frac{1}{m}\sum_{i=1}^m \rho_i^{\otimes t} = \int_{\Omega_N}\dd{\rho}_\text{HS} \rho^{\otimes t}.
	\end{equation}
	Certain examples of mixed-state $t$-designs in $\Omega_N$ can be obtained from vectors forming a complex projective $t$-designs in ${\cal H}_N \otimes {\cal H}_N$ by partial trace. Examples of such a construction described in \cite{CGGZ20} include a $2$-design given by the complete set of $5$ mutually unbiased bases of order $4$ with all $20$ vectors sharing the same degree of entanglement.
	
	\subsection{Helstrom theorem}
	
	Orthogonal quantum states can be perfectly distinguished in a single measurement. In general, to characterize distinguishability between two arbitrary states $\rho$ and $\sigma$ one uses their trace distance, $D_\text{tr}(\rho,\sigma) = \frac{1}{2}\Tr\abs{\rho - \sigma}$, where the prefactor $\frac{1}{2}$ ensures that the diameter of the set of states is fixed to unity, $D_\text{tr} \qty(\op{0},\op{1}) = 1$. The single-shot distinguishability between them is described by the celebrated Helstrom theorem \cite{H69}.
	
	\begin{theorem}{\textbf{(Helstrom)}}
		Given two arbitrary quantum states $\rho, \sigma \in \Omega_N$, the probability $p$ of distinguishing between them in a single-shot experiment, if both occur with equal probability, is upper bound by 
		\begin{equation} \label{eq:hell_theor}
			p \leq \frac{1}{2}\qty(1 + D_\text{tr}(\rho,\sigma)).
		\end{equation}
		The bound is saturated by the measurement based on the eigenvectors of the difference of the quantum states, $\rho - \sigma=\sum_i \lambda_i \op{i}$, and given by measurement operators
		
		\begin{align}
			P_\rho = \sum_{\lambda_i > 0} \lambda_i \op{ i}, &&
			P_\sigma = \sum_{\lambda_i < 0} \lambda_i \op{i}, &&
			P_0 = \mathbb{I} - P_\rho - P_\sigma.
		\end{align}
	\end{theorem}
	
	In this work the single-sided distinguishability is understood as distinguishability in the sense of the Helstrom theorem between single-party reductions of states belonging to the composite Hibert space, $\rho_j= {\rm Tr_B}\op{\psi^{AB}}$. Local distinguishability, which we do not consider here, takes into account the use of local operations and classical communication (LOCC) between parties \cite{WSHV00, Co07, ZGQZW15}. Single-party distinguishability, as used by Akibue and Kato \cite{AK18}, refers to treating a bipartite system as a whole single party. In this sense, the states we are seeking are perfectly distinguishable as they form an orthonormal basis. However, a pair of two-qubit states is single-party distinguishable if and only if both states are pure and separable, $\ket{\psi_1} = \ket{\phi^A_1}\otimes\ket{\phi^B_1}$ and $\ket{\psi_2} = \ket{\phi^A_2}\otimes\ket{\phi^B_2}$, with orthogonal states in the both subsystems, $\braket{\phi_1^A}{\phi_2^A} = 0 = \braket{\phi_1^B}{\phi_2^B}$.
	
	\subsection{Elegant Joint Measurement}

	Let us recall the question posed by Massar and Popescu \cite{MP95} concerning the optimal extraction of information from a finite ensemble of identically polarized spin-$1/2$ particles. In order to solve this problem they formulated a game in which a player, Alice, receives a certain number of such spins, parallel to each other and oriented in a particular well-defined direction in the space, random in each run of the game. She is allowed to make a measurement in any global basis. Taking into account these results she guesses a particular direction of the spins. The guess must be deterministic, without probabilities for different directions. The score of each run is $\cos[2](\alpha/2)$, where $\alpha$ is the angle between the guessed direction and the actual one. The score of the whole game is the average of scores of all runs. The authors identified and described an optimal two-qubit measurement, which gives the maximal winning chances.
	
	An analogous configuration, related to the SIC measurement superposed with the singlet Bell state was analysed later by Gisin \cite{G17}, who introduced the name {\sl Elegant Joint Measurement} (EJM). This particular measurement scheme was recently applied to study a three-party quantum network \cite{G19}. It was demonstrated that the probabilities $p_{\tr}(a,b,c)$ of getting results $a$, $b$ and $c$ by Alice, Bob and Charlie respectively, generated by EJM, display a high degree of correlation. This level of correlation cannot be generated by any three-local model, thus providing an example of a non-three-local quantum triangle. In the subsequent section we will show that in the case $N=2$ this generalized measurement leads to a two-qubit orthogonal basis with an optimal single-sided distinguishability. 
	
	\section[Bipartite bases in $\mathcal{H}_N \otimes {\cal H}_N$ with maximal single-sided distinguishability]{Bipartite bases in $\mathcal{H}_N \otimes {\cal H}_N$ 
		\\with maximal single-sided distinguishability} \label{sec:bases}
	
	\subsection{Basis with optimal one-party distinguishability for two-qubit system}
	
	We wish to identify a collection of $N^2$ pure states $\qty{\ket{\psi_i}}_{i=1}^{N^2}$ comprising an orthonormal basis on the space $\mathcal{H}_N^{\otimes 2}={\cal H}_A \otimes {\cal H}_B$, for which the single-sided distinguishability between any two states is optimal. Such a basis is characterized by the maximal distance between all reduced states, $\rho^A_i = \Tr_B\op{\psi_i}$ and $\rho^B_i = \Tr_A\op{\psi_i}$, in each subsystem, where $i=1,\dots, N^2$. More precisely, we require that the trace distance between all reductions is constant for all $i \ne j$, and that the constant $L$ is as large as it can be,
	
	\begin{equation}\label{eq:dmax}
		D_\text{max} = \underset{\qty{\ket{\psi_i}}}{\max}\qty{L:\, \forall_{i \ne j}\,D_\text{tr}\qty(\rho^A_i,\rho^A_j) = D_\text{tr}\qty(\rho^B_i,\rho^B_j) = L}.
	\end{equation}
	Imposing equality of distances between any two single-party reduced states leads to two simplices of side $D_\text{max}$ with respect to the trace distance embedded inside the set of mixed states of dimension $N$, the same for both subsystems. 
	
	This simplicial property is easy to visualize in the simplest example for a pair of qubits, $N=2$. Consider a set of four pure states that constitute a SIC generalized measurement for a single-qubit system, which form a tetrahedron inscribed in the Bloch sphere,
	
	\begin{align*}
		\ket{1_0} = \mqty(1 \\ 0), && 
		\ket{2_0} = \frac{1}{\sqrt{3}}\mqty(1 \\ \sqrt{2}), &&
		\ket{3_0} = \frac{1}{\sqrt{3}}\mqty(1 \\ \omega\sqrt{2}), && 
		\ket{4_0} = \frac{1}{\sqrt{3}}\mqty(1 \\ \omega^{2}\sqrt{2})
	\end{align*}
	with $\omega = \exp(i\frac{2\pi}{3})$. It is also convenient to introduce a dual SIC configuration, distinguished by the orthogonality relation, $\braket{i_0}{i_1} = 0$, which forms the dual tetrahedron,
	
	\begin{align*}
		\ket{1_1} = \mqty(0 \\ 1), && 
		\ket{2_1} = \frac{1}{\sqrt{3}}\mqty(\sqrt{2} \\ -1), && 
		\ket{3_1} = \frac{1}{\sqrt{3}}\mqty(\sqrt{2} \\ -\omega), && 
		\ket{4_1} = \frac{1}{\sqrt{3}}\mqty(\sqrt{2} \\ -\omega^{2}).
	\end{align*}
	Both SIC schemes together yield a total of $2^3$ states comprising the compound of SIC measurements recently studied by Tavakoli et al. \cite{TBGR20}, which can be divided into two separate SIC configurations or four separate orthonormal basis.
	
	We shall also use the notion of a conjugate state defined by the complex conjugation of all components of the state in the computational basis. Such conjugated states will be denoted by $\ket{i_j^*}$ for both SIC measurements.
	
	The local states forming both SIC configurations allow us to introduce a family of four bipartite states forming locally tetrahedral structures in both single-party subspaces,
	\begin{align}
		\ket{\psi_i} & = \sqrt{\lambda}\ket{i_0}\otimes\ket{i^*_0} + e^{i \phi_i}\sqrt{1 - \lambda}\ket{i_1}\otimes\ket{i_1^*}, \label{eq:qb1}
	\end{align}
	with the Schmidt coefficient $\lambda \in [0,1]$ and phases $\phi_i \in [0,2\pi]$ treated as free parameters. In what follows we will be interested in optimizing the Schmidt coefficient $\lambda$ defining the degree of entanglement of the states, which in turn maximizes the trace distance between the reduced states, $D_\text{tr}\qty(\rho^A_i,\,\rho^A_j)$.
	
	At the same time we impose orthogonality restrictions on these states in order to form an orthonormal basis in the total space $\mathcal{H}_{AB} = \mathcal{H}_A \otimes {\cal H}_B$,
	\begin{align}
		\braket{\psi_i}{\psi_j} = \frac{1}{3}\qty(\lambda + e^{i(\phi_j - \phi_i)}\qty(1-\lambda) + 2\sqrt{\lambda\qty(1-\lambda)}\qty(e^{-i \phi_i} + e^{i \phi_j})) = 0.
	\end{align}
	
\begin{figure}[H]
		\centering
		\includegraphics[width=\linewidth]{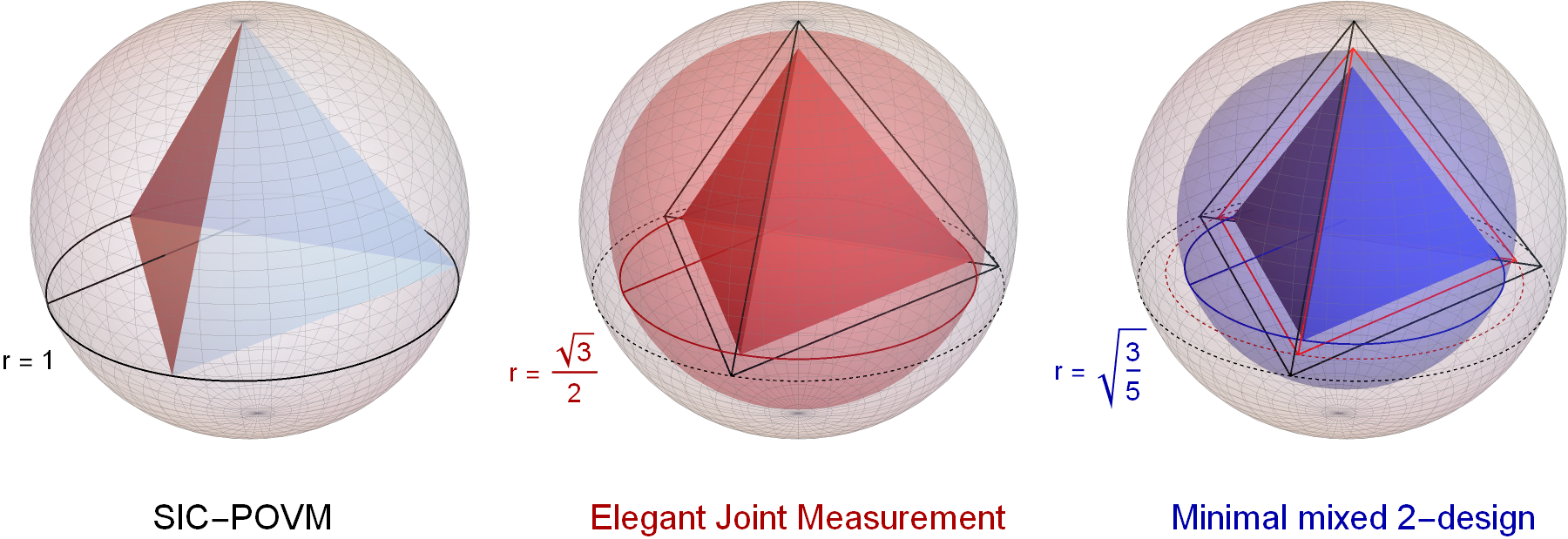}
		\caption{The orthonormal basis for 2 qubits with maximal single-sided distinguishability $\qty{\ket{\psi_i}}$ induces by reduction a tetrahedral structure, depicted in blue, inscribed in a sphere of radius $r = \sqrt{3}/2$ inside the Bloch sphere, lying between the larger tetrahedron of SIC measurement (black), of radius $r_\text{SIC} = 1$, and the minimal example of mixed-state 2-design, making up the smaller tetrahedron of radius $r_\text{mix} = \sqrt{3/5}$ \cite{CGGZ20}. Note that the radii of three spheres form the series  $\qty{\sqrt{\frac{3}{3}},\sqrt{\frac{3}{4}},\sqrt{\frac{3}{5}}}$
			decreasing as $x^{-1/2}$.}
		\label{fig:simplex_comparison}
	\end{figure}

	Any two such restrictions for two pairs of indices $i,k$ and $j,k$ imply equality of the phases, $\phi_i = \phi_j \equiv \phi$. Next, by simple algebraic manipulation we arrive at an equation connecting the phase $\phi$ with the Schmidt coefficient $\lambda$,
	
	\begin{equation}
		\cos\phi = \frac{-1}{4\sqrt{\lambda(1 - \lambda)}},
	\end{equation}
	which implies that $\lambda$ satisfies the inequality
	
	\begin{equation}\label{eq:2_lambda_ineq}
		\frac{1}{16} \leq \lambda(1 - \lambda).
	\end{equation}
	This relation allows us to deduce that the maximal basis with respect to the single-sided distinguishability is found by setting the phase to zero $\phi = 0$ and maximizing the Schmidt coefficient, $\lambda_{\normalfont\text{max}} = \frac{1}{2}\qty(1 + \frac{\sqrt{3}}{2})$. The basis obtained in this way can be written explicitly row-wise as a unitary matrix
	
		\begin{equation} \label{eq:EJM_unitary}
			U_4 = \mqty(\bra{\psi_1} \\ \bra{\psi_2} \\ \bra{\psi_3} \\ \bra{\psi_4}) = \mqty(
			\frac{1}{2}\sqrt{2 + \sqrt{3}} & 0 & 0 & -\frac{1}{2}\sqrt{2 - \sqrt{3}} \\
			\frac{1}{6} \sqrt{6-3 \sqrt{3}} & \frac{1}{\sqrt{3}} & \frac{1}{\sqrt{3}} & \frac{1}{6} \sqrt{6+3 \sqrt{3}} \\
			\frac{1}{6} \sqrt{6-3 \sqrt{3}} & \frac{\omega^2}{\sqrt{3}} & \frac{\omega}{\sqrt{3}} & \frac{1}{6} \sqrt{6+3 \sqrt{3}} \\
			\frac{1}{6} \sqrt{6-3 \sqrt{3}} & \frac{\omega}{\sqrt{3}} & \frac{\omega^2}{\sqrt{3}} & \frac{1}{6} \sqrt{6+3 \sqrt{3}}
			),
	\end{equation}
	the order of its columns can be set arbitrarily. It is important to note that the optimal basis found for $N=2$ reproduces the Elegant Joint Measurement (EJM) obtained by Gisin \cite{G19}. To compare this particular collection of states with already known tetrahedral configurations including the standard single-qubit SIC measurement and the minimal mixed-state 2-design \cite{CGGZ20}, let us discuss some common measures of entanglement of bi-partite pure states. The values of the purity $\gamma$, linear entropy $S_L$, von Neumann entropy $S$ and the Schmidt angle $\theta_S$, characterizing the entanglement of the pure states $\ket{\psi_i}$ forming the optimal 2-qubit basis and the degree of mixing of the reduced states $\rho_i = \Tr_B \op{\psi_i}$ read 
	\begin{align}
		\gamma(\rho_i) & = \Tr\qty(\rho_i^2) 
		= \lambda_{\normalfont\text{max}}^2 + \qty(1 - \lambda_{\normalfont\text{max}})^2 = \frac{7}{8},
		&
		S_L(\rho_i) & \equiv 1 - \gamma(\rho_i)
		= \frac{1}{8}, 
		\nonumber\\
		S(\rho_i) & \equiv -\Tr(\rho_i \log \rho_i) \approx 0.246, &
		\theta_S(\rho_i) & \equiv \arcsin \lambda_{\normalfont\text{max}} \approx 1.2027.
	\end{align}
	As all the states forming the bases are iso-entangled, the above values are the same for all the states of the basis and do not depend on the index $i$. The states carry an intermediate entanglement, as they are neither maximally entangled nor product, thus they contain both local and global information. Comparing the mean purity of reduced states with the other examples presented in Fig.\ref{fig:simplex_comparison} we find that
	
	\begin{align}
		\ev{\Tr(\rho^2)} & = 
		\begin{cases}
			1 & \text{for SIC measurements} \\
			\frac{7}{8} & \text{for elegant joint measurements} \\
			\frac{4}{5} & \text{for mixed--state 2-designs}.
		\end{cases}
	\end{align}
	Observe that the EJM tetrahedron is located between the other two tetrahedra, as the partial states are more pure than those forming the minimal mixed-state 2-design, but not entirely pure as the states in the one-qubit SIC measurement. It is worth noting that the radii of the spheres for the three constructions considered form an interesting series, $\qty{\sqrt{\frac{3}{3}},\sqrt{\frac{3}{4}},\sqrt{\frac{3}{5}}}$.
	
	Moreover, one can investigate the properties of EJM as a whole, interpreting it as a unitary matrix. It is known \cite{KC01} that any two-qubit unitary gate $U \in SU(4)$ can be decomposed as
	\begin{equation}
		U = \qty(V_1 \otimes V_2)U_c\qty(W_1 \otimes W_2)
	\end{equation}
	where $V_1,\,V_2,\,W_1,\,W_2\in U(2)$ are local rotations, corresponding to preparation and post-processing. The nonlocal two-qubit gate $U_c$ can be represented in the Cartan form,
	
	\begin{equation} \label{eq:cartan_deco}
		U_c = \exp \qty(  i\sum_{i=1}^3\varphi_i \sigma_i\otimes\sigma_i)
	\end{equation}
	with phases $\varphi_i\in\mathbb{R}$ and $\sigma_i$ being the Pauli matrices. The coordinates of EJM in terms of the Cartan phases $\phi_i$ are given by $\qty{\pi/4, \pi/8,0}$, which are exactly the same as for the $B$--gate introduced by Zhang \textit{et al.} \cite{ZVSW04}, which has been shown to allow implementation of any arbitrary 2-qubit unitary gate in a minimal number of 1- and 2-qubit gates. 
	
\begin{figure}[H]
		\centering
		\includegraphics[width=.9\linewidth]{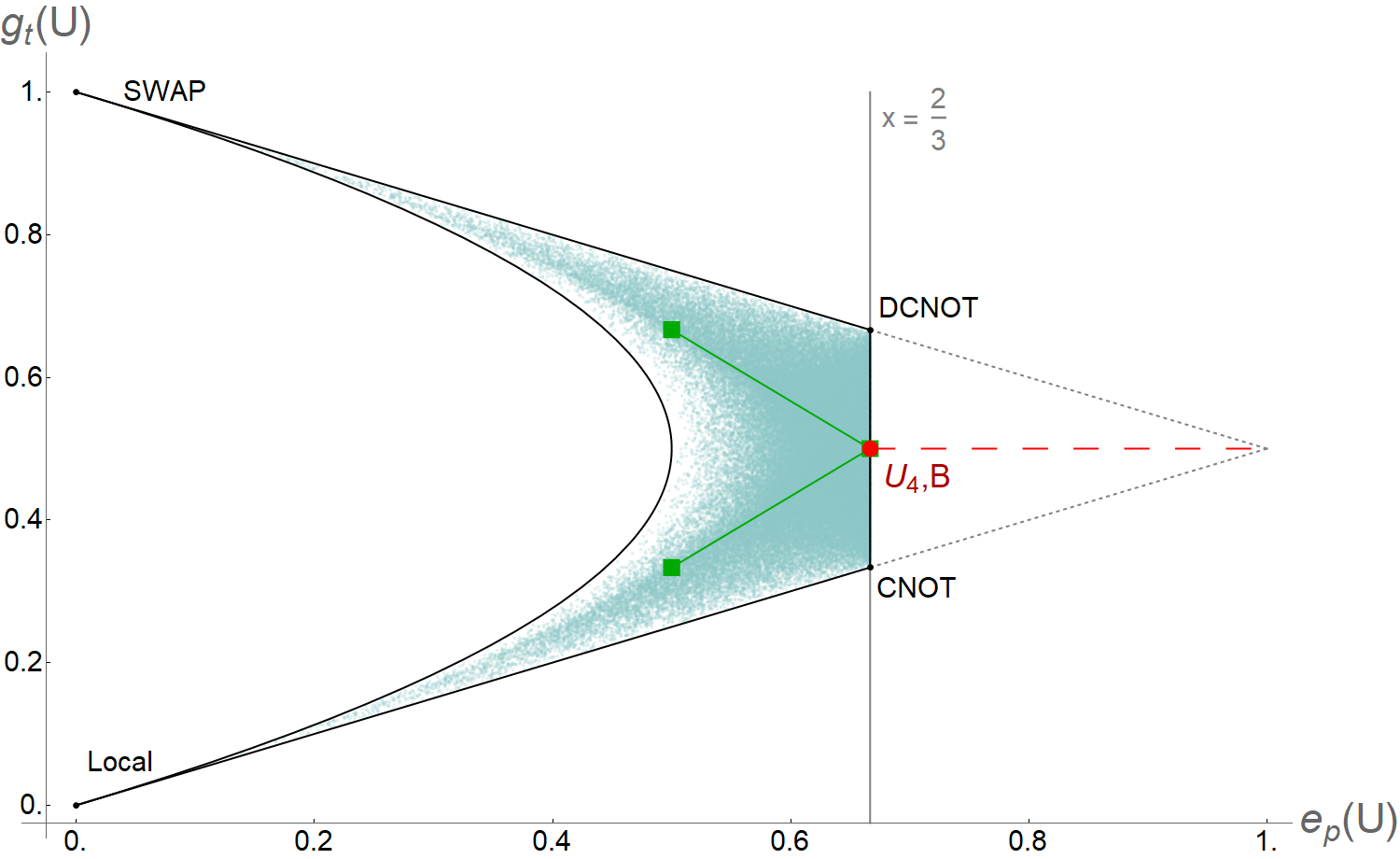}
		\caption{The Elegant Joint Measurement, characterized by the matrix $U_4$, yields a single point (in red) in terms of entangling power $e_p(U_4)$ and gate typicality $g_t(U_4)$ with coordinates $\qty(2/3,1/2)$ independently of the order of the vectors. This point lies on the boundary of allowed set for 2-qubit operations, as shown by the thick black line, and is the closest one to the (unattainable) 2-unitary gate with coordinates $\qty(1,1/2)$. It can be compared with the $B$--gate introduced by Zhang \textit{et al.} \cite{ZVSW04}, which gives exactly the same point. Permuting rows of the $B$--gate, one gets unitary matrices represented by points $(1/2,1/3)$ and $(1/2,2/3)$, given in dark green. The teal points filling the black contour have been generated from a sample of $10^5$ random unitary matrices of dimension 4.}
		\label{fig:ep_gt}
	\end{figure}

	Yet another way of characterizing a unitary operation is its entangling power $e_p(U)$ and 
	the complementary quantity, gate typicality $g_t(U)$, as defined in \cite{JMZL20} by the operator operator linear entropy $E(U) = 1 - \frac{1}{N^4} \sum_{i=1}^{N^2} \lambda_i^2$ with $\lambda_i$ being the eigenvalues of the reshuffled matrix $U^R$,
	
	\begin{align}
		e_p(U) = \frac{E(U) + E(US) - E(S)}{E(S)}, &&
		g_t(U) = \frac{E(U) - E(US) + E(S)}{2E(S)}
	\end{align}
	where $S$ is the SWAP operator.	We find that $e_p(U_4) = \frac{2}{3}$ and, due to symmetry $U_4S = \overline{U_4}$, $g_t(U_4) = \frac{1}{2}$ for arbitrary order of the basis vectors. When compared to the possible range of these values for 2-qubit operations, as in Fig. \ref{fig:ep_gt}, it is easily seen that the entangling power is maximal, whereas the gate typicality is exactly in the middle. The resulting point is closest possible for 2-qubit gates to a 2-unitary gate \cite{GALRZ15} and in this sense is again found to be closely related to the $B$--gate studied by Zhang \textit{et al.} \cite{ZVSW04}.
	
	\subsection{Orthogonal basis with optimal single-sided distinguishability for $N>2$}
	
	After exhibiting the simplest, two-qubit case, we can proceed to the main result of this work. It consists in providing a constructive proof that orthogonal bases with desired properties can be found in higher dimensions for any $N$, for which a SIC measurement is known.

	\begin{theorem}\label{teo:lambdamax}
	Consider a composite bipartite space $\mathcal{H}_N^{\otimes 2} = \mathcal{H}^A_N\otimes\mathcal{H}^B_N$ and a given set of vectors
	$\qty{\ket{i_0}\in\mathcal{H}_N}_{i=1}^{N^2}$, forming a full SIC measurements on the corresponding subspaces $\mathcal{H}_N$. There always exists a family of orthonormal bases $\qty{\ket{\psi_i}}_{i = 1}^{N^2}$ on the composite space, for which single-party, mutual distinguishability between all $N^2$ states is identical for any $i\neq j$. The basis reads
	\begin{equation}
		\label{eq:optbfase_prop}
		\ket{\psi_i} = \sqrt{\lambda}\ket{i_0}\ket{i_0^*} - \sqrt{\frac{1 - \lambda}{N-1}}\sum_{j=1}^{N-1} \ket{i_j}\ket{i_j^*},
	\end{equation}
	where the states $\ket{i_j}\in\mathcal{H}_N^A$ belong to the orthogonal spaces of SIC vectors satisfying relations $\braket{i_j}{i_k} = \delta_{jk}$. Vectors $\ket{i_j^*}\in\mathcal{H}_N^B$ are obtained by conjugating components of $\ket{i_j}$ for $i=1,\hdots,N^2$ and $j = 0,\,\hdots,\,N-1$ . 
	
	Within this family the Schmidt coefficient $\lambda_{\normalfont\text{max}}$ providing maximal single-sided distinguishability is given as
	
	\begin{equation} \label{eq:lambdamax}
		\lambda_{\normalfont\text{max}} = \frac{N^3-N^2-N+2 (N-1) \sqrt{N+1}+2}{N^3}.
	\end{equation}
	\end{theorem}
	A proof by construction of such an optimal bases provided in Appendix \ref{app:teo_lmax} works for any dimension $N$, for which a SIC construction is known. The corresponding Schmidt vectors are given by $\Lambda = \mqty(\lambda_{\normalfont\text{max}}, & \lambda, & \hdots &, \lambda)$ with $\lambda = \frac{1 - \lambda_{\normalfont\text{max}}}{N - 1}$. An alternative expression  of the state  (\ref{eq:optbfase_prop}), involving the maximally entangled state, is provided in Eq.  (\ref{super_entan}).

	In order to evaluate the single-sided distinguishability we represent the reduced states by density matrices,
	\begin{align}
		\rho_i & = \Tr_B\qty(\op{\psi_i}) = \frac{N\lambda_{\normalfont\text{max}} - 1}{N-1}\op{i_0} + \frac{N\qty(1 - \lambda_{\normalfont\text{max}})}{N-1} \rho_*
	\end{align}
	where $\rho_* = \mathbb{I}/N$ denotes the maximally mixed state in dimension $N$. By this token, the trace distance between these states is easily calculated, 
	\begin{align}\label{eq:dmax_explicit}
		D_\text{max}(N) = D_\text{tr}\qty(\rho_i,\rho_j)
		& = \qty(1 - \delta_{ij})\frac{\qty(N-2)\sqrt{N+1} +2}{N^{3/2}}.
	\end{align}
	and we may conjecture that this value is the optimal among all bases.
	
	\begin{conjecture}\label{conj:optimality}
		The basis given in the Theorem \ref{teo:lambdamax} is optimal among all possible bases in $\mathcal{H}_N^{\otimes 2}$ for which $D_{\normalfont{\text{tr}}}(\rho_i,\rho_j) = \normalfont{\text{const.}}$
	\end{conjecture}
	
	Let us now formulate the following observation concerning the limit of a large dimension.
	
	\begin{remark} \label{rem:asym}
		For a large dimension $N$ the largest Schmidt coefficient corresponding to the optimal basis approaches unity,
		\begin{equation}
			\lim_{N\rightarrow\infty} \lambda_{\normalfont\text{max}} = 1.
		\end{equation}
	\end{remark}
	This implies that in the limit of large dimension $N$ the reduced states $\Tr_A \op{\psi_i}$ and $\Tr_B \op{\psi_i}$ converge to the corresponding pure SIC vectors, which provide an optimal local tomography scheme, as it forms a complex projective 2-design \cite{S06}. 
	
	Since existence of SIC measurements was conjectured for every dimension $N$ \cite{Z99, Z11}, we put forward conjectures applicable in particular to the dimensions $N$, for which the SIC problem is not yet settled. 
	\begin{conjecture}\label{conj:exist}
		For any integer $N$ there exists an orthonormal basis $\qty{\ket{\psi_i}}_{i =1}^{N^2}$ on the composite bipartite space $\mathcal{H}_N \otimes \mathcal{H}_N$, such that the partial traces $\qty{\rho_i = \Tr_B \op{\psi_i}}$ and $\qty{\sigma_i = \Tr_A \op{\psi_i}}$ form maximal simplices of side length $D_\text{max}$, given by Eq.\eqref{eq:dmax_explicit}, according to the trace distance on the corresponding spaces $\Omega_N$,
		\begin{equation}
			\forall_{i,j}:\,
			D_\text{tr}\qty(\rho_i,\,\rho_j) = 
			D_\text{tr}\qty(\sigma_i,\,\sigma_j) = 
			D_\text{\normalfont\text{max}}.
		\end{equation}
	\end{conjecture}
	
	\begin{conjecture}\label{conj:equiv}
		Existence of a basis $\qty{\ket{\psi_i}}$ postulated in Conjecture \ref{conj:exist} on the space $\mathcal{H}_N^{\otimes 2}$ is equivalent to the existence of a SIC measurement, determined by the eigenvectors $\ket{i}$ of states $\rho_i$ corresponding to the leading eigenvalue.
	\end{conjecture}
	Implication in one direction follows from Theorem \ref{teo:lambdamax}, where a construction of the basis $\qty{\ket{\psi_i}\in\mathcal{H}_{N^2}}$ from a SIC configuration in $\mathcal{H}_N$ is derived. The reverse implication remains open.

	\section{Elegant Joint Measurement for $N=3$} \label{sec:EJM3}
	
	In order to show the proposed construction in action we present an explicit solution of the problem for $N=3$. As a first step let us consider a SIC-POVM for dimension $N = 3$, given by $9$ equiangular vectors
	
	\begin{align}\label{eq:q3_sic_states}
		\ket{1_0} = \ket{1_0^*} & = \mqty(1\\0\\0), &
		\ket{2_0} = \ket{3_0^*} & = \frac{1}{2}\mqty(1 \\ i\sqrt{3} \\ 0), &
		\ket{3_0} = \ket{2_0^*} & = \frac{1}{2}\mqty(1 \\ -i\sqrt{3} \\ 0), \nonumber \\
		\ket{4_0} = \ket{4_0^*} & = \frac{1}{2}\mqty(1 \\ 1 \\ \sqrt{2}), & 
		\ket{5_0} = \ket{6_0^*} & = \frac{1}{2}\mqty(1 \\ 1 \\ \omega\sqrt{2}), & 
		\ket{6_0} = \ket{5_0^*} & = \frac{1}{2}\mqty(1 \\ 1 \\ \omega^2\sqrt{2}), \nonumber\\ 
		\ket{7_0} = \ket{7_0^*} & = \frac{1}{2}\mqty(1 \\ -1 \\ \sqrt{2}), &
		\ket{8_0} = \ket{9_0^*} & = \frac{1}{2}\mqty(1 \\ -1 \\ \omega\sqrt{2}), & 
		\ket{9_0} = \ket{8_0^*} & = \frac{1}{2}\mqty(1 \\ -1 \\ \omega^2\sqrt{2}). 
	\end{align}
	with $\omega = \exp(i \frac{2\pi}{3})$. By applying Theorem \ref{teo:lambdamax} we easily find the global basis with optimal single-sided distinguishability for two qutrits:
	
	\begin{proposition} \label{prop:q3_basis}
		An orthonormal basis with optimal single-sided distinguishability on the space $\mathcal{H}_3^{\otimes 2}$ can be given as a single unitary matrix formed row-wise by the state vectors,	

	\begin{equation}
	\label{U9}
	U_9 = 		{\scriptsize\mqty(
		\bra{\psi_1} \\
		\bra{\psi_2} \\
		\bra{\psi_3} \\
		\bra{\psi_4} \\
		\bra{\psi_5} \\
		\bra{\psi_6} \\
		\bra{\psi_7} \\
		\bra{\psi_8} \\
		\bra{\psi_9}) 
		= \frac{1}{6\sqrt{6}}\mqty(
		10 \sqrt{2} & 0 & 0 & 0 & -2 \sqrt{2} & 0 & 0 & 0 & -2 \sqrt{2} \\
		\sqrt{2} & -3 i \sqrt{6} & 0 & 3 i \sqrt{6} & 7 \sqrt{2} & 0 & 0 & 0 & -2 \sqrt{2} \\
		\sqrt{2} & 3 i \sqrt{6} & 0 & -3 i \sqrt{6} & 7 \sqrt{2} & 0 & 0 & 0 & -2 \sqrt{2} \\
		\sqrt{2} & 3 \sqrt{2} & 6 & 3 \sqrt{2} & \sqrt{2} & 6 & 6 & 6 & 4 \sqrt{2} \\
		\sqrt{2} & -3 \sqrt{2} & 6 & -3 \sqrt{2} & \sqrt{2} & -6 & 6 & -6 & 4 \sqrt{2} \\
		\sqrt{2} & 3 \sqrt{2} & 6 \omega & 3 \sqrt{2} & \sqrt{2} & 6 \omega & 6 \omega^2 & 6 \omega^2 & 4 \sqrt{2} \\
		\sqrt{2} & 3 \sqrt{2} & 6 \omega^2 & 3 \sqrt{2} & \sqrt{2} & 6 \omega^2 & 6 \omega & 6 \omega & 4 \sqrt{2} \\
		\sqrt{2} & -3 \sqrt{2} & 6 \omega & -3 \sqrt{2} & \sqrt{2} & -6 \omega & 6 \omega^2 & -6 \omega^2 & 4 \sqrt{2} \\
		\sqrt{2} & -3 \sqrt{2} & 6 \omega^2 & -3 \sqrt{2} & \sqrt{2} & -6 \omega^2 & 6 \omega & -6 \omega & 4 \sqrt{2} \\
		)}
	\end{equation}
		with $\omega = \exp(2 i \pi/3)$.
	\end{proposition}

\begin{figure}[H]
		\centering
		\includegraphics[width=\linewidth]{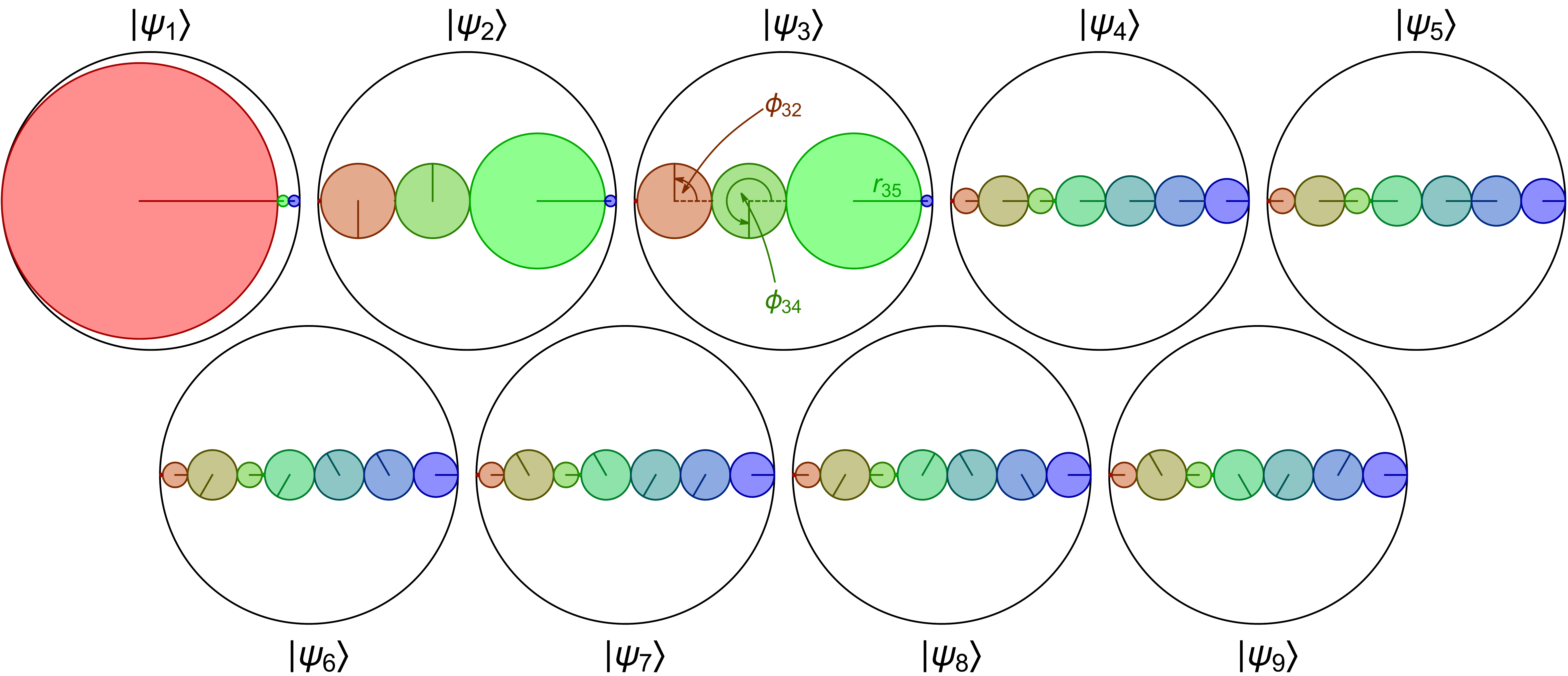}
		\caption{Each vector $\ket{\psi_i} = \sum_{j=1}^9 a_{ij}\ket{j}$ from the basis given by $U_9$ in \eqref{U9} can be visually represented by a set of 9 colour-labelled clocks with with a single hand each. The radii $r_{ij}$ of the clocks are given by the squared modulus of the respective components, so that $\sum_{j=1}^9 r_{ij} = \sum_{j=1}^9 \abs{a_{ij}}^2 = 1$, as indicated by the outer black circles. The angles between the $x$ axis and the indicators are given by the arguments of the components, $\varphi_{ij} = \text{arg}\qty(a_{ij})$.}
		\label{fig:qutrit_vis}
	\end{figure}

	The basis $U_9$ is visualized in Fig. \ref{fig:qutrit_vis} by the clock representation of the nine constituting states.
	
	Note that this explicit two-qutrit measurement can be considered as a generalization of the Elegant Joint Measurement for $N = 3$. Such a basis may find its use in a triangular network setting and possibly non-trilocal correlations for parties sharing qutrits instead of qubits.
 	Furthermore, a more symmetric solution $U_{9\text{sym}}$ based on a different form of SIC-POVM is presented in the Appendix \ref{app:symm}. Additionally, entangling power and gate typicality of the bases $U_9$, $U_{9\text{sym}}$ and the maximally entangled basis $U'_9$ given in Appendix \ref{app:maxmix} are presented in Appendix \ref{app:ep_gt}.
	
	\section{Tomographical scheme and its experimental implementation} \label{sec:experiment}

	\subsection{Tomographical capability of bases with maximal distinguishability}
	
	We are going to demonstrate that the orthogonal bases in bi-partite systems introduced earlier in this work are more than just a mathematical curiosity and may find practical applications concerning quantum state tomography. First, by the Zauner conjecture we assume that a SIC generalized measurement exists for every dimension $N$ \cite{Z99, Z11}, assuring existence of the basis with maximal single-sided distinguishability in dimension $N^2$. We can now ask whether this basis can provide us with a tomographical scheme with a useful reconstruction formula. Consider a scheme based on $N^2$ states from Theorem \ref{teo:lambdamax} that gives by partial trace a rescaled SIC measurement in dimension $N$, 
	\begin{equation}
		\rho_i = \frac{N \lambda_{\normalfont\text{max}} - 1}{N - 1}\op{i_0} + \frac{1 - \lambda_{\normalfont\text{max}}}{N - 1}
		\mathbb{I}.
	\end{equation}
	By simple algebraic manipulation of two-copy states $\tilde{\rho}_i^{\otimes2}$ of properly rescaled states $\tilde{\rho}_i = \frac{1}{N}\rho_i$, so that $\sum \tilde{\rho}_i = \mathbb{I},$ we arrive at a reconstruction formula of an analysed state $\sigma\in\Omega_N$, akin to the one provided by the SIC,
	
	\begin{equation}
		\sigma = \frac{(N-1)^2 N (N+1)}{(\lambda_{\normalfont\text{max}} N-1)^2}\sum_{i=1}^{N^2}\Tr(\tilde{\rho}_i\sigma)\tilde{\rho}_i - \frac{2 \lambda_{\normalfont\text{max}} +N^2-\left(\lambda_{\normalfont\text{max}} ^2+1\right) N-1}{(\lambda_{\normalfont\text{max}} N-1)^2}\mathbb{I}
	\end{equation}
	where the corresponding measurement probabilities read $p_i = \Tr(\tilde{\rho}_i\sigma)$.
	
	In the single-qubit case, $N=2$, we find an explicit reconstruction formula for the Elegant Joint Measurement \cite{MP95},
	\begin{equation} \label{eq:qubit_case_reco}
		\sigma = 8\sum_{i=1}^4 p_i \tilde{\rho}_i - \frac{3}{2}\mathbb{I}
	\end{equation}
	with probabilities of getting a given outcome, $p_i = \Tr(\sigma\tilde{\rho}_i)$. It is natural to expect certain noise in the entire measurement procedure, resulting in probabilities inferred up to an accuracy, $p'_i = p_i + \Delta p_i$. By plugging them into the equation \eqref{eq:qubit_case_reco} we may quantify the departure of the reconstructed state from the real state,
	
	\begin{equation}
		\Delta \sigma_{\text{EJM}} = 8\sum_{i=1}^4 \Delta p_i \tilde{\rho}_i.
	\end{equation}
	Similar considerations for a SIC measurement performed on a single system yields
	\begin{equation}
		\Delta \sigma_{\text{SIC}} = 6\sum_{i=1}^4 \Delta p_i \tilde{\rho}_i.
	\end{equation}
	Comparison between the two estimations of errors clearly shows that the tomographical procedure obtained from EJM is not much inferior to SIC-POVM. Better still, the advantage of a SIC shrinks with dimension $N$ as the reductions of the bases introduced in this work approach SIC asymptotically, as stated in Remark \ref{rem:asym}. One of the key features of the scheme advocated in this work is the possibility to use the bipartite orthogonal basis to measure simultaneously both subsystems equally well. This allows for gathering the information concerning both parties independently at two experimental sites using the same setup and  implementing the corresponding bipartite L{\"u}ders-von Neumann measurement. 
	
	\subsection{Experimental implementation of the proposed tomography scheme}
	
	In order to demonstrate the tomographical capability of the proposed measurement scheme on real devices we treat the case $N=2$, equivalent to the Elegant Joint Measurement. In the first step we need to translate it into a quantum circuit. To achieve this task we follow an approach suitable for any two-qubit gates, as opposed to an earlier scheme of a circuit corresponding to EJM proposed in \cite{BGT20}. 
%
	Any two-qubit gate in the Cartan form \eqref{eq:cartan_deco} can be decomposed in terms of two-qubit gates $CNOT_{ij}$, (acting on the $j$-th qubit with the control qubit labelled by $i$), and local operations \cite{VW04}
	\begin{equation}
		U_c = CNOT_{21} \qty[R_z(\gamma)\otimes R_y(\alpha)] CNOT_{12} \qty(\mathbb{I}\otimes R_y(\beta))CNOT_{21}.
	\end{equation}
	with $\alpha, \beta, \gamma$ expressable in terms of the information content of the gate $U_c$ as
	
	\begin{align*}
		\alpha = 2\phi_3 - \frac{\pi}{2} && \beta = \frac{\pi}{2} - 2\phi_1 && \gamma = 2\phi_2 - \frac{\pi}{2}
	\end{align*}
	
\begin{figure}[H]
		\centering
		\includegraphics[width=0.75\linewidth]{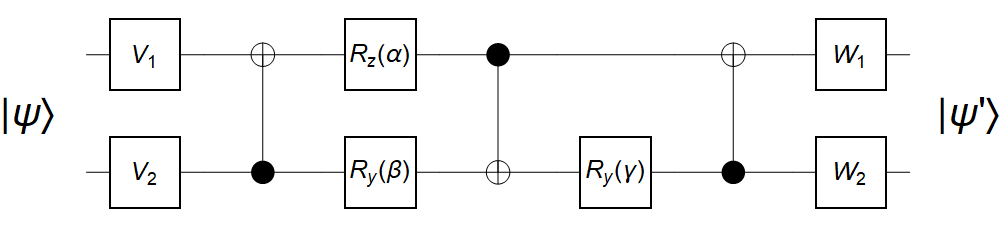}
		\caption{General scheme for decomposition of any $U\in SU(4)$, using local pre- and post-preparation local gates $V_{1,2},W_{1,2}$, and entangling subcircuit \cite{VW04}. Such circuits should be read from left to right, so that the entire scheme gives $\ket{\psi'} = U \ket{\psi}$, as composed stepwise starting from the local operations $V_1\otimes V_2$ followed by the two-qubit CNOT gate.}
		\label{fig:circ_gen}
	\end{figure}

	Collecting all ingredients we arrive at the circuit representation of the gate $U_c$, schematically depicted in Fig. \ref{fig:circ_gen}.
	
	After deriving the circuit corresponding to the EJM, described in detail in Appendix \ref{app:circ}, the next step is to consider how to implement the single-party measurements determined by mixed states. Consider a POVM composed of operators proportional to rank-1 projectors $\qty{\Pi_i}$, which satisfies the resolution of identity, $\sum \Pi_i = \mathbb{I}$. The probability that the measurement of state $\rho$ will return the $i$-th result reads,
	\begin{equation}
		p_i = \Tr(\rho \Pi_i).
	\end{equation}
	Let us now suppose that $\rho \in \Omega_{N^2}$, so that it can be seen as a bipartite state. Moreover, assume it has a tensor product form, $\rho = \frac{\mathbb{I}_A}{N} \otimes \sigma_B \equiv \frac{\mathbb{I}}{N} \otimes \sigma$, where we drop subscripts denoting Alice and Bob for later convenience. For such a state the probability is given by
	
	\begin{equation}
		p_i = \Tr[\qty(\frac{\mathbb{I}}{N} \otimes \sigma)\Pi_i] = 
		\Tr[\sigma \frac{\Tr_A\qty(\Pi_i)}{N}].
	\end{equation}		
	
	\begin{figure}[h]
		\centering
		\includegraphics[width=\linewidth]{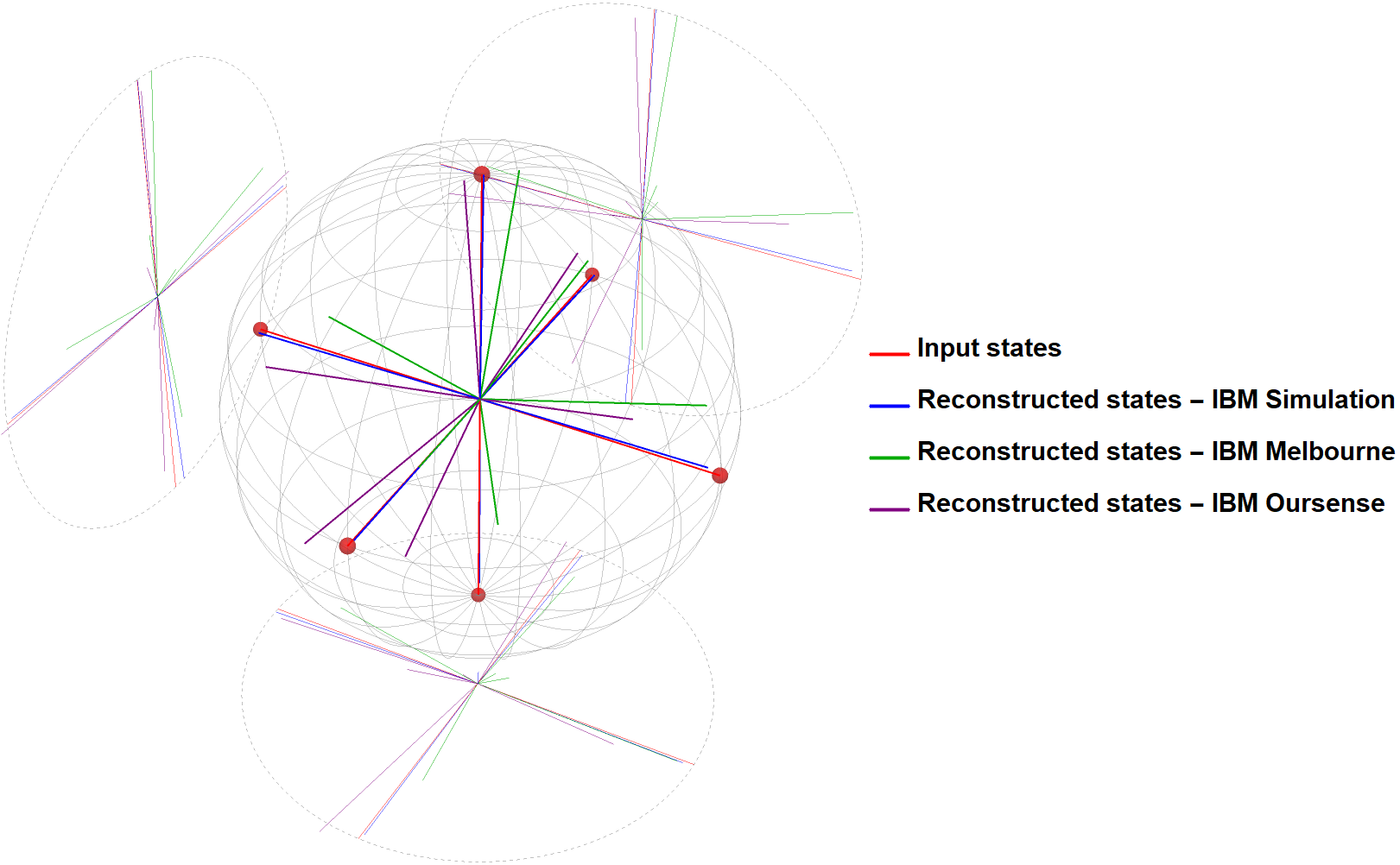}
		\caption{Reconstruction of one-qubit states constituting a full set of MUBs (red pure states) for a qubit, using the scheme with auxiliary Bell pair, depicted in a Bloch ball with additional projections on $XY$, $XZ$ and $YZ$ for more clarity. Blue states are reconstructed from simulated measurements on a quantum computer. Violet and green states have been measured on real-world
			quantum devices - IBM Oursense 5-qubit T-architecture computer and IBM Melbourne 15-qubit ladder-architecture computer, respectively.}
		\label{fig:reco_maxmix}
	\end{figure}
	
	By performing such a measurement, we implement a POVM consisting of operators proportional to partial reductions of the original POVM. It can be effected on the system by adding an auxiliary qubit to a system C, which is entangled with system A, thus having $\Tr_C\qty(\op{\Psi_+}_{AC}\otimes \sigma_B) = \frac{\mathbb{I}_A}{2} \otimes \sigma_B$, as intended for the implementation of our desired POVM.  Here $\ket{{\Psi_+}_{AC}}=(\ket{0_A0_C} +\ket{1_A 1_C})/\sqrt{2}$ denotes the Bell state shared between subsystems $A$ and $C$. 
	
	\medskip
	
	We have tested this scheme in single runs on Melbourne and Oursense quantum computers provided by IBM, with $8192$ shots for each state reconstruction. After taking into account the success rates, described in Appendix \ref{app:calib}, the estimated effective sizes of samples are $6608$ for Melbourne and $7551$ for the Oursense machine. Results of this measurement are visualized in Fig. \ref{fig:reco_maxmix}.
	
	The first step was to simulate the expected results using IBM's cloud-based simulation. This yielded reconstruction almost identical to the original MUBs, as simulations  do not take into account realistic  levels of noise and imperfections thus giving overly optimistic estimations. Next, the experiments were taken to real-world computers, IBM Melbourne and Oursense quantum computing machines. Here we can clearly see the effects of errors both on gates and qubits, which can be checked in the calibration data provided in Appendix \ref{app:calib}. Taking into account the estimated success rates, given in Table \ref{tab:calibs}, that are on the level of 80\% for Melbourne machine and 92\% for Oursense, the reconstruction results turn out to be satisfactory.

	\section{Concluding Remarks}
	\label{sec:conclude}

	The search for optimal schemes of a quantum measurement applicable for various set-ups is an ongoing task in the field of quantum information. In this work we investigated the question of finding an orthogonal von Neumann measurement for two parties having $N$ levels each, which provides the optimal single-sided distinguishability in terms of single-shot experiments. Such particular orthogonal bases of size $N^2$ induce, by the partial trace, simplicial structures in both subspaces. The optimal mutual distinguishability between reduced states is ensured by the largest size of the $N^2-1$ dimensional simplex embedded inside the set $\Omega_N$ of mixed states of order $N$.
	
	The above question is solved as we provide a general construction of an optimal bases in the composite space $\mathcal{H}_N \otimes \mathcal{H}_N$, which implies the maximal single-sided distinguishability between any two reduced density matrices. The presented construction is valid  in every dimension $N$, in which a SIC measurement is known. In this way we extend the connection between the L{\"u}ders -- von Neumann orthogonal measurement on bipartite systems and single party simplex-like structures for a given number of $N$ local levels. All $N^2$ states forming the basis are iso-entangled, and their entanglement is quantified by the size of the maximal component $\lambda_{\normalfont\text{max}}$ of the corresponding Schmidt vector.
	
	In the case of $N=2$, corresponding to the two-qubit system, the solution found reproduces the configuration of states in $\mathcal{H}_4$ called the Elegant Joint Measurements introduced by Gisin  \cite{G17,G19}. The related basis \eqref{eq:EJM_unitary} forms a 2-qubit gate distinguished by the maximal entangling power $e_p = 2/3$ and gate typicality $g_t = 1/2$, making it locally equivalent to the $B$--gate, optimal with respect to construction of arbitrary 2-qubit gates with a minimal number of 1- and 2-qubit gates \cite{ZVSW04}. 
	
	The qutrit basis  \eqref{U9} in $\mathcal{H}_9$ can be thus considered as a generalization of EJM for the case $N=3$. Furthermore, the unitary matrix $U'_9$ corresponding to the maximally entangled basis for qutrits is equivalent, up to a permutation, to a 2-unitary matrix \eqref{eq:max_ep_qutrits}, maximizing the entangling power, $e_p = q$. In the limit of the large dimension $N$ the reduced states become perfectly distinguishable and tend asymptotically to the pure states corresponding to a SIC measurement in each subspace. We conjecture that existence of an orthogonal basis in $\mathcal{H}_N \otimes \mathcal{H}_N$ saturating the upper bound for single-sided distinguishability implies existence of SIC measurement in the corresponding single party space $\mathcal{H}_N$. 
	
	Furthermore, we demonstrate the tomographical capability of introduced bases in a single-party reduction. The obtained tomographical scheme is evaluated explicitly for qubit systems and compared to SIC measurements, showing minor decrease in precision, which vanishes asymptotically as the local dimension grows. The proposed scheme for quantum state tomography for a qubit system related to the two-qubit optimal orthogonal basis was implemented on IBM quantum computers.  Our results demonstrate experimental capability of the introduced tomographic procedure.
	
	There are several open questions and numerous ways in which results of this work can be extended. Let us mention here a natural problem of finding distinguished bases in a composite Hilbert space $\mathcal{H}_N^{\otimes m}$ describing an $m$-party system, such that the one-sided distinguishability between all single-system reduced states is maximal. Already for $m=3$ subsystems, one can additionally ask about the maximal distinguishability among states composed of $m-1$ subsystems, if only a single subsystem is traced away. Another question concerns a possible generalization of the scheme of Massar and Popescu \cite{MP95} for the case of particles with a higher spin.

	\section*{Acknowledgments}

	It is a pleasure to thank N. Gisin for helpful remarks that allowed us to introduce additional simplification to final expressions. Moreover, we are obliged to  I. Bengtsson, D.~Goyeneche, M. Grassl, K. Korzekwa, A. Lakshminarayan, Z. Pucha{\l}a,   O. Reardon-Smith and A. Tavakoli  for inspiring discussions and valuable correspondence. Financial support by Narodowe Centrum Nauki under the grant numbers DEC-2015/18/A/ST2/00274 and 2019/35/O/ST2/01049 and by Foundation for Polish Science under the Team-Net NTQC project is gratefully acknowledged.
	
	We acknowledge use of the IBM Q for this work. The views expressed are those of the authors and do not reflect the official policy or position of IBM or the IBM Q team.
	
	\appendix
	\section{Proof of Theorem \ref{teo:lambdamax}} \label{app:teo_lmax}
	
	As a starting point, let us consider a single-party $N$-dimensional Hilbert space $\mathcal{H}_N$ and suppose we have two states $\ket{\vb{x}_0}$ and $\ket{\vb{y}_0}$ such that their inner product satisfies the condition imposed on states within a SIC-POVM,
	
	\begin{equation}
		\abs{\braket{\vb{x}_0}{\vb{y}_0}}^2 = \braket{\vb{x}_0}{\vb{y}_0}\braket{\vb{x}_0^*}{\vb{y}_0^*} = \frac{1}{N+1}.
	\end{equation}
	
	For both of these vectors we define perpendicular spaces $S_{\vb{x}} = \text{span}\qty(\qty{\ket{\vb{x}_i}}_{i=1}^{N-1})$ and similarly for $\ket{\vb{y}}$ and conjugate vectors, where the states $\ket{\vb{x}_i}$ and $\ket{\vb{y}_i}$ can be chosen in such a way that they satisfy the relation
	$			
	\braket{\vb{x}_i}{\vb{x}_j} = 
	\braket{\vb{y}_i}{\vb{y}_j} = 
	\braket{\vb{x}_i^*}{\vb{x}_j^*} = 
	\braket{\vb{y}_i^*}{\vb{y}_j^*} = \delta_{ij}.
	$
	for $i, j = 0,\hdots,N-1$.
	
	We can now proceed to the two-party isoentangled states of interest $\ket{\psi_{\vb{x}}},\ket{\psi_{\vb{y}}}\in \mathcal{H}_N^{\otimes 2}$. We will construct them based on the given two SIC states under the natural assumption that \textbf{no direction in the perpendicular space $S_i$ is distinguished}. This leads to an \textit{Ansatz} in which all components related to the perpendicular spaces will have the same Schmidt coefficients, so that the bi-partite states from $\mathcal{H}^{\otimes 2}_N$ can be expressed as
	
	\begin{align}
		\ket{\psi_{\vb{x}}} = \sqrt{\lambda}\ket{\vb{x}_0}\ket{\vb{x}_0^*} + e^{i\phi} \sqrt{\frac{1 - \lambda}{N-1}}\sum_{i=1}^{N-1} \ket{\vb{x}_i}\ket{\vb{x}_i^*}, \\
		\ket{\psi_{\vb{y}}} = \sqrt{\lambda}\ket{\vb{y}_0}\ket{\vb{y}_0^*} + e^{i\phi} \sqrt{\frac{1 - \lambda}{N-1}}\sum_{i=1}^{N-1} \ket{\vb{y}_i}\ket{\vb{y}_i^*}.
	\end{align}
	In this way we arrive at states which will form the first two vertices of a regular simplex in terms of trace distance $D_{\text{tr}}$ in both reductions to a single party. The corresponding reductions to the first party read
	
	\begin{align}
		\rho_{\vb{x}}  & = \frac{N\lambda - 1}{N - 1} \op{\vb{x}_0} + \frac{1 - \lambda}{N - 1} \mathbb{I}, &
		\rho_{\vb{y}}  & = \frac{N\lambda - 1}{N - 1} \op{\vb{y}_0} + \frac{1 - \lambda}{N - 1} \mathbb{I}
	\end{align}
	and the trace distance between them is given by
	
	\begin{equation}
		D_{\text{tr}} = \frac{N\lambda - 1}{2\qty(N-1)}\norm{\op{\vb{x}_0} - \op{\vb{y}_0}}_1.
	\end{equation}
	Hence this value becomes maximal for the maximal admissible value of $\lambda$.
	
	We will now impose the orthogonality relation, so that these two orthogonal vectors will give two elements of the desired orthonormal basis in $\mathcal{H}_N \otimes \mathcal{H}_N$. Thus, we calculate the inner product explicitly,
	\begin{align}
		\braket{\psi_{\vb{x}}}{\psi_{\vb{y}}} = & 
		\lambda \braket{\vb{x}_0}{\vb{y}_0}\braket{\vb{x}_0^*}{\vb{y}_0^*} + 
		\frac{1-\lambda}{N-1}\sum_{i,j}^{N-1}\braket{\vb{x}_i}{\vb{y}_j}\braket{\vb{x}_i^*}{\vb{y}_j^*} = 0
	\end{align}
	and impose the orthogonality condition. This can be expressed in a simple form, $A\lambda + B + C\sqrt{\lambda - \lambda^2}=0$, with coefficients $A, B, C$ introduced for brevity,
	\begin{align}
		A = \frac{N(N-2)}{N^2-1}, && 
		B = \frac{N^2 - N - 1}{N^2 - 1}, && 
		C = \frac{2 N \cos (\phi )}{\qty(N+1)\sqrt{N-1}}.
	\end{align}

	The above equation can be easily solved, 
	\begin{equation}
		\lambda_\pm = \frac{C^2 - 2AB \pm \sqrt{\Delta}}{2(A^2 + C^2)}
	\end{equation}
	with $\Delta = C^2\qty(C^2 - 4B^2 - 4AB)$. The solution is valid for $C < 0$, so that the signs of $C\sqrt{\lambda - \lambda^2}$ and $A\lambda + B$ have to be opposite.
	
	Since we are looking for bases with maximal distinguishability of the partial traces, we are interested in maximizing $\lambda$ under the constraint, that the vectors are orthogonal. It is easily found, that the only parameter we control is the phase $\phi$, and the maximum (as well as minimum) $\lambda$ corresponds to the maximum of $C^2$ and, in turn, to the maximum of $\cos^2 \phi$, which implies $\phi = \pi$. For such a choice of the phases the bounds for the Schmidt coefficients $\lambda_{\normalfont\text{max}}$ and $\lambda_\text{min}$ read
	
	\begin{align}
		\lambda_{\normalfont\text{max}} & = \frac{N^3-N^2-N+2 (N-1) \sqrt{N+1}+2}{N^3} \label{eq:app_lambdamax}\\
		\lambda_\text{min} & = -\frac{-N^3+N^2+N+2 (N-1) \sqrt{N+1}-2}{N^3}. \label{eq:app_lambdamin}
	\end{align}
	The upper value $\lambda_{\normalfont\text{max}}$ yields the desired bound claimed in Theorem \ref{teo:lambdamax}. \qed
	
	As a final remark, we note that the bipartite pure states forming an optimal basis can be written using a maximally entangled generalized  Bell state, $\ket{\psi_+} = \frac{1}{\sqrt{N}}\sum_{j=1}^N \ket{j,j}$. Observe that the following relation holds,
	\begin{equation}
		\sum_{i=0}^{N-1} \ket{\vb{x}_i}\ket{\vb{x}^*_i} = \sum_{j=1}^N \ket{j,j} \equiv \sqrt{N}\ket{\psi_+},
	\end{equation}
	where $\ket{j,j}$ are to be understood as vectors from the computational basis of the bipartite $N^2$-dimensional space. Thus any basis state can be considered as a coherent superposition of  a product state and the maximally entangled state,	
	\begin{equation}
		\label{super_entan}
		\ket{\psi_{\vb{x}}} = \qty(\sqrt{\lambda_\text{max}}+\sqrt{\frac{1 - \lambda_{\text{max}}}{N-1}})\ket{i_0}\ket{i_0^*} - \sqrt{\frac{N(1 - \lambda_{\text{max}})}{N-1}}\ket{\psi_+}.
	\end{equation}
	Observe the similarity between the above expression and the form of the two-qubit EJM originally constructed by  Gisin \cite{G17,G19}.
	
	\section{Maximally entangled solutions for $N = 2$ and $3$} \label{app:maxmix}
	
	It is interesting to note that Theorem \ref{teo:lambdamax} deals with the maximal case, but in the course of derivation one could in principle introduce single-parameter families of bases stemming from SIC measurements in corresponding dimension, with varying degree of entanglement of the states. In particular, the upper and lower bounds for $\lambda$ coefficient for qubits, $N=2$, calculated in \eqref{eq:2_lambda_ineq} allow for a maximally entangled solution, $\lambda = 1/2$ and $\phi = \frac{\pi}{3}$, which we show below, arranged as the rows of a unitary matrix $U'_{4}$,

	\begin{equation}
		U'_{4} = \mqty(
		\bra{\psi'_1} \\
		\bra{\psi'_2} \\
		\bra{\psi'_3} \\
		\bra{\psi'_4}) = \frac{1}{3\sqrt{2}}\mqty(
		\sqrt{6} & 0 & 0 & \sqrt{6}\omega_6^4 \\
		\sqrt{2} & 2\omega_6^2 & 2\omega_6^2 & \sqrt{2}\omega_6 \\
		\sqrt{2} & 2 & 2\omega_6^4 & \sqrt{2}\omega_6 \\
		\sqrt{2} & 2\omega_6^4 & 2 & \sqrt{2}\omega_6 \\
		),
	\end{equation}
	with $\omega_6 = \exp(2\pi i/6)$. In this case the underlying tetrahedron degenerates to a point,\linebreak $\Tr_A\op{\psi_i} = \mathbb{I}/2$. This solution does not lead to any single-sided distinguishability, as all partial traces are maximally mixed. However, after such a measurement both parties have perfect correlations between the states measured, apart from the mirror symmetry implied by the complex conjugation of states between them. This reproduces one of the extreme points of the family introduced by Tavakoli et al. in \cite{TGB20}. Interestingly, when interpreted as a unitary operation, $U'_4$ has the same entangling power $e_p$, gate typicality $g_t$ and information content $\phi_i$ as the matrix $U_4$ introduced in Eq. \eqref{eq:EJM_unitary}.
	
\begin{figure}[ht!]
		\centering
		\includegraphics[width=\linewidth]{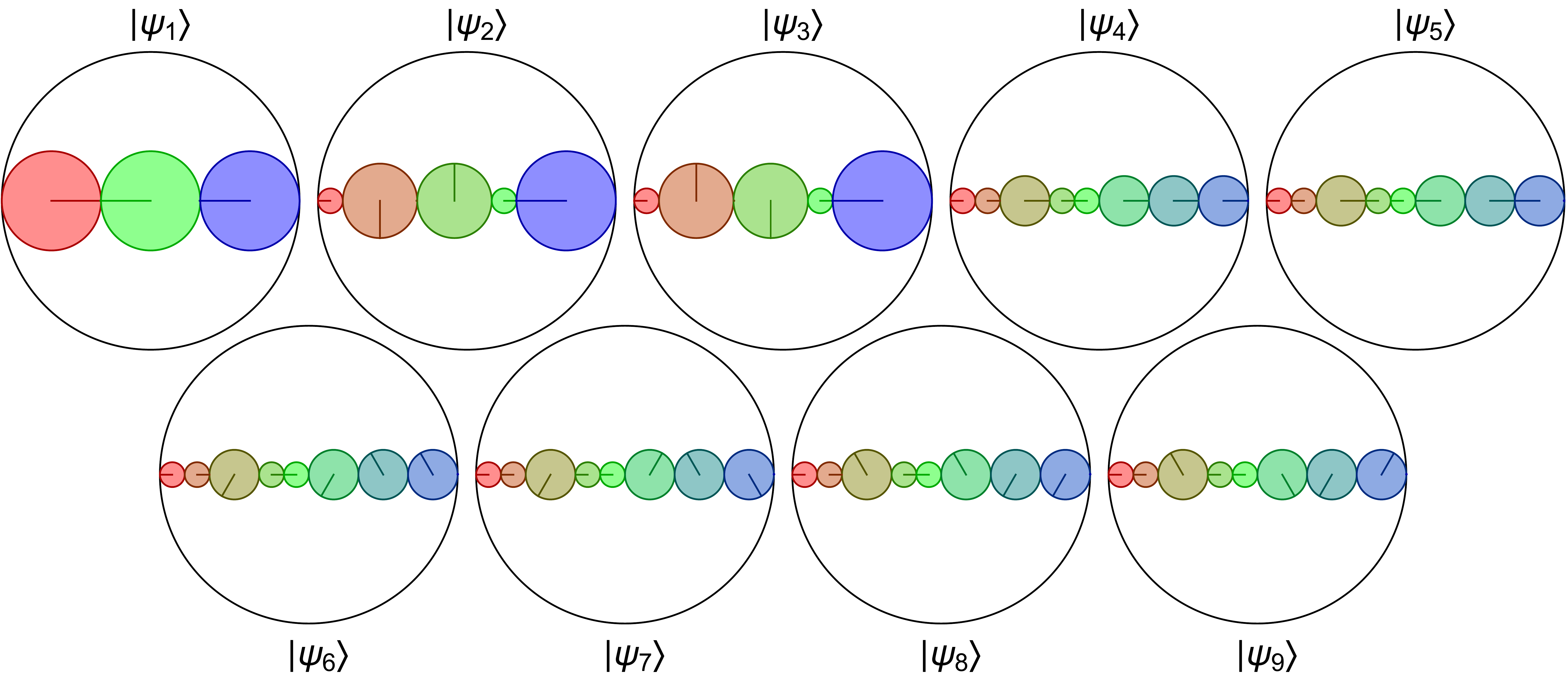}
		\caption{Clock for basis $U'_9$ from equation \eqref{u9maxent}, constructed as in Fig. \ref{fig:qutrit_vis}.}
		\label{fig:maxent_qutrit_vis}
	\end{figure}

	Similarly, the minimal solution for $N = 3$ is particularly compelling, as we find it to have $\lambda = 1/3$, which corresponds to the maximally mixed states. This is impossible for any higher dimensions, since one can easily check that $\lambda_{\text{min}} > 1/N$ for $N > 3$. This solution, arranged below in a unitary matrix $U'_{9}$, is worth noticing for this particular reason,
	
	\begin{equation}\label{u9maxent}
		U'_{9} = \frac{1}{2\sqrt{3}}\mqty(
		\ 2 & 0 & 0 & 0 & -2 \ & 0 & 0 & 0 & -2 \ \  \\
		-1 & -i \sqrt{3} & 0 & i \sqrt{3} &\ \  1 & 0 & 0 & 0 & -2\  \ \\
		-1 & i \sqrt{3} & 0 & -i \sqrt{3} &\  \ 1 & 0 & 0 & 0 & -2\  \ \\
		-1 & 1 & \sqrt{2} & 1 & -1 & \sqrt{2} & \sqrt{2} & \sqrt{2} & 0 \\
		-1 & -1\ \   & \sqrt{2} & -1\ \  & -1 & -\sqrt{2} & \sqrt{2} & -\sqrt{2} & 0 \\
		-1 & 1 & \omega^2 \sqrt{2} & 1 & -1 & \omega^2 \sqrt{2} & \omega \sqrt{2} & \omega \sqrt{2} & 0 \\
		-1 & -1\ \  & \omega^2 \sqrt{2} & -1\ \ & -1 & -\omega^2 \sqrt{2} & \omega \sqrt{2} & -\omega \sqrt{2} & 0 \\
		-1 & 1 & \omega \sqrt{2} & 1 & -1 & \omega \sqrt{2} & \omega^2 \sqrt{2} & \omega^2 \sqrt{2} & 0 \\
		-1 & -1\ \  & \omega \sqrt{2} & -1 \ \ & -1 & -\omega \sqrt{2} & \omega^2 \sqrt{2} & -\omega^2 \sqrt{2} & 0
		)
	\end{equation}
	can be visualized using the clock representation introduced for $U_9$ -- see Fig. \ref{fig:maxent_qutrit_vis}.
		
	\section{Symmetric solution for $N = 3$} \label{app:symm}
	
	The basis $U_9$ given in \eqref{U9} proceeds from the simplest, most widely used form of SIC-POVM for $N=3$. 
	
\begin{figure}[ht!]
		\centering
		\includegraphics[width=.7\linewidth]{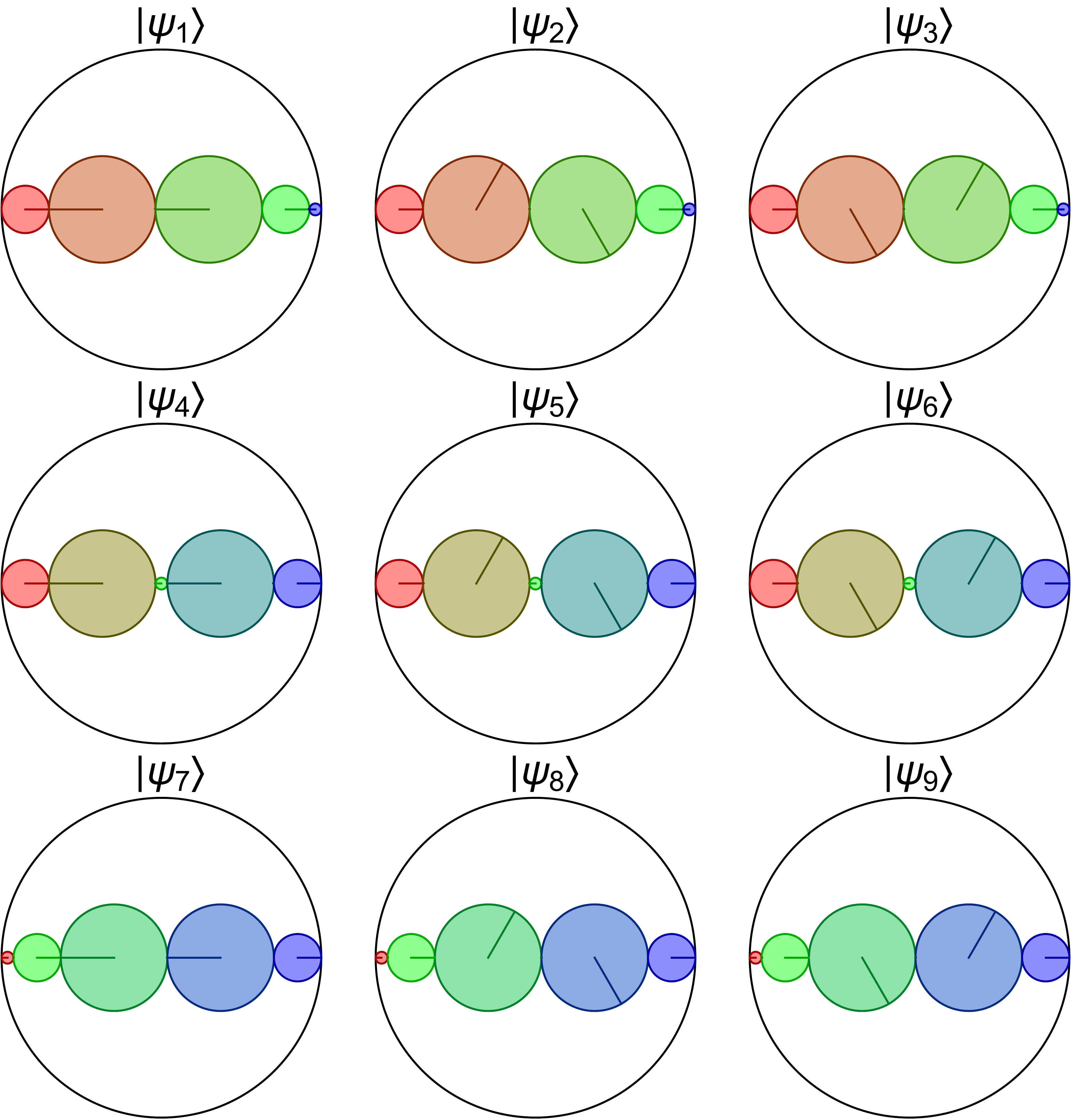}
		\caption{Clock representation for basis $U_{9\text{sym}}$, constructed as in Fig. \ref{fig:qutrit_vis}.}
		\label{fig:qutritvisualizationsymmetric}
	\end{figure}

	One can use a more symmetric form of SIC-POVM, given by
	
	\begin{align}
		\ket{1_0} = \ket{1_0^*} & = \frac{1}{\sqrt{2}}\mqty(1 \\  -1 \\ 0), &
		\ket{2_0} = \ket{3_0^*} & = \frac{1}{\sqrt{2}}\mqty(1 \\ -\omega \\ 0), &
		\ket{3_0} = \ket{2_0^*} & = \frac{1}{\sqrt{2}}\mqty(1 \\ -\omega^2 \\ 0), \\
		\ket{4_0} = \ket{4_0^*} & = \frac{1}{\sqrt{2}}\mqty(1 \\ 0 \\ -1), &
		\ket{5_0} = \ket{6_0^*} & = \frac{1}{\sqrt{2}}\mqty(1 \\ 0 \\ -\omega), &
		\ket{6_0} = \ket{5_0^*} & = \frac{1}{\sqrt{2}}\mqty(1 \\ 0 \\ -\omega^2), \\
		\ket{7_0} = \ket{7_0^*} & = \frac{1}{\sqrt{2}}\mqty(0 \\ 1 \\ -1), &
		\ket{8_0} = \ket{9_0^*} & = \frac{1}{\sqrt{2}}\mqty(0 \\ 1 \\ -\omega), &
		\ket{9_0} = \ket{8_0^*} & = \frac{1}{\sqrt{2}}\mqty(0 \\ 1 \\ -\omega^2)
	\end{align}
	with $\omega = \exp{i \frac{2\pi}{3}}$, to arrive at a basis revealing a significant level of symmetry,

		\begin{equation} \label{U9sym}
			U_{9\text{sym}} = \frac{1}{3\sqrt{3}}\mqty(
			-1 & 0 & 0 & 0 & 2 & -3 & 0 & -3 & 2 \\
			-1 & 0 & 0 & 0 & 2 & -3 \omega^2 & 0 & -3 \omega & 2 \\
			-1 & 0 & 0 & 0 & 2 & -3 \omega & 0 & -3 \omega^2 & 2 \\
			2 & 0 & -3 & 0 & -1 & 0 & -3 & 0 & 2 \\
			2 & 0 & -3 \omega^2 & 0 & -1 & 0 & -3 \omega & 0 & 2 \\
			2 & 0 & -3 \omega & 0 & -1 & 0 & -3 \omega^2 & 0 & 2 \\
			2 & -3 & 0 & -3 & 2 & 0 & 0 & 0 & -1 \\
			2 & -3 \omega^2 & 0 & -3 \omega & 2 & 0 & 0 & 0 & -1 \\
			2 & -3 \omega & 0 & -3 \omega^2 & 2 & 0 & 0 & 0 & -1
			).
		\end{equation}
	
	directly reflected in Fig. \ref{fig:qutritvisualizationsymmetric}.
	
	Similarly, starting from this symmetric SIC-POVM one can derive a maximally entangled basis,
	
	\begin{equation}
		E'_{9\text{sym}} = \frac{1}{\sqrt{3}}\mqty(
			0 & 0 & 1 & 0 & 1 & 0 & 1 & 0 & 0 \\
			0 & 0 & \omega^2 & 0 & 1 & 0 & \omega & 0 & 0 \\
			0 & 0 & \omega & 0 & 1 & 0 & \omega^2 & 0 & 0 \\
			0 & 1 & 0 & 1 & 0 & 0 & 0 & 0 & 1 \\
			0 & \omega^2 & 0 & \omega & 0 & 0 & 0 & 0 & 1 \\
			0 & \omega & 0 & \omega^2 & 0 & 0 & 0 & 0 & 1 \\
			1 & 0 & 0 & 0 & 0 & 1 & 0 & 1 & 0 \\
			1 & 0 & 0 & 0 & 0 & \omega^2 & 0 & \omega & 0 \\
			1 & 0 & 0 & 0 & 0 & \omega & 0 & \omega^2 & 0).
	\end{equation}

	\section{Entangling power and gate typicality for $N=3$ solutions} \label{app:ep_gt}
	
	It is simple to analyse the bases $U_9$, $U_{9\text{sym}}$ and $U'_9$ in terms of their entangling power $e_p$ and gate typicality $g_t$ \cite{JMZL20}. 
	
\begin{figure}[ht!]
		\centering
		\includegraphics[width=.8\linewidth]{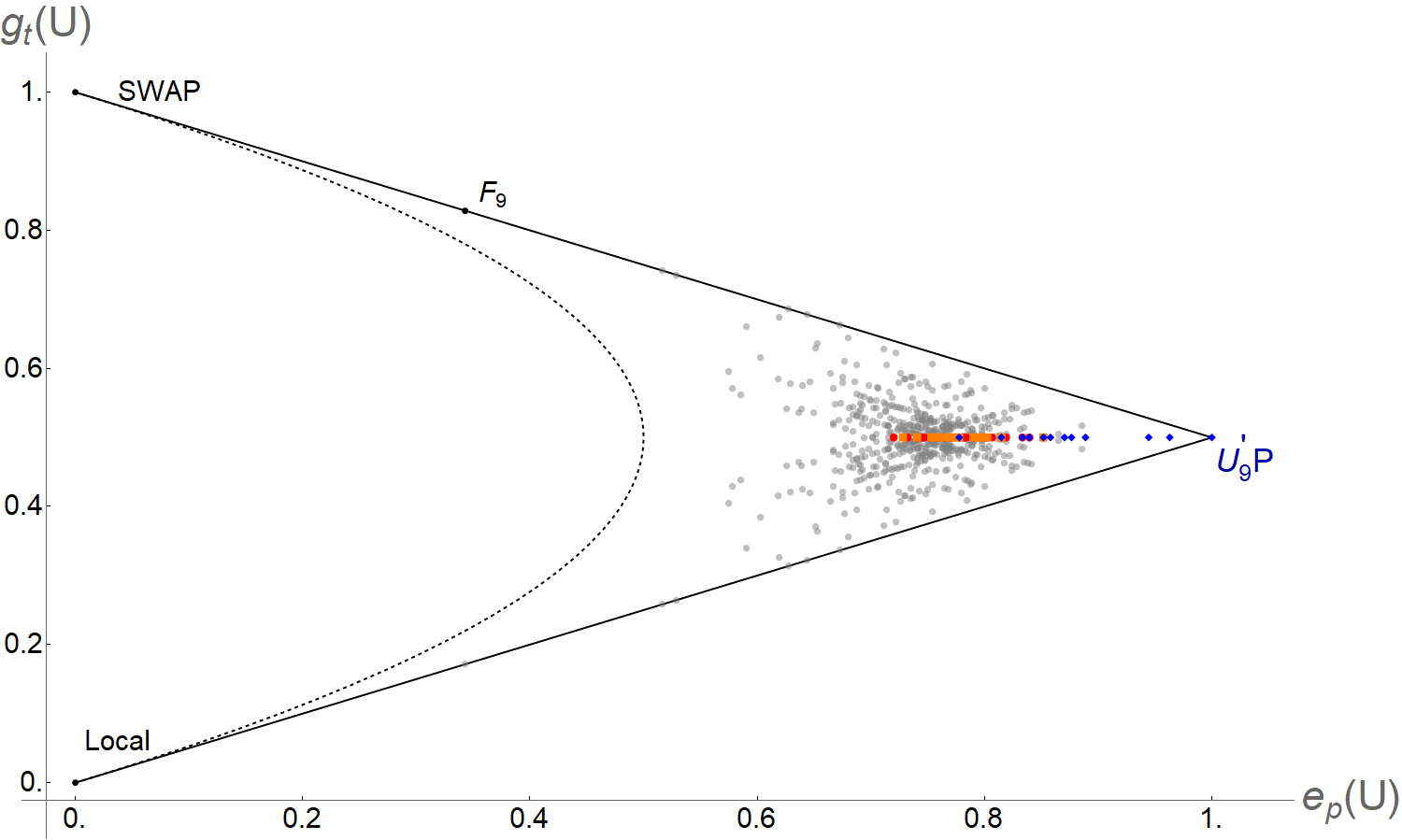}
		\caption{The qutrit bases $U_9$ (red), $U_{9\text{sym}}$ (orange) and $U'_9$ (blue) can be visualized in the plane
			of entangling power and gate typicality, where the left boundary of the allowed region can be approximated by the parabola, with single points found to the ``west'' of it \cite{JMZL20}. Similarly to the $U_4$ case, we consider all possible orders of the states in the basis, which in the case of qutrits turns out to be important, as we find that depending on the permutation the entangling power $e_p$ can vary, even though $g_t = 1/2$ is common across all possibilities due to the symmetry $US = \overline{U}$. In particular, we find that for maximally entangled basis $U'_9$ there exists a subset of permutations for which the operation the maximal entangling power $e_p = 1$ and thus is 2-unitary \cite{GALRZ15}. The gray points in the background correspond to all $543$ points obtained from the permutations of the Fourier matrix $F_9$.}
		\label{fig:ep_gt_qtrit}
	\end{figure}

	As we have done with the EJM, here we analyze all possible orders of the vectors in the basis, which generically would give raise to $9! \approx 3*10^5$ distinct points. For our basis we find only 24, 12 and 12 distinct points, respectively. For comparison, the Fourier matrix, defined by $\qty(F_9)_{jk} = \exp(2 i \pi \frac{j\cdot k}{9})$ gives rise to a total of 543 distinct points in the $\qty(e_p, g_t)$ plane, as shown in Fig. \ref{fig:ep_gt_qtrit}. 
	
	Important feature of all the introduced bases is that their gate typicality is exactly halfway, ie. $g_t = \frac{1}{2}$. Moreover, a certain permutation $P$ applied to the maximally entangled basis $U'_9$,
	
	\begin{equation}\label{eq:max_ep_qutrits}
		{\scriptsize U_9' P = \frac{1}{2\sqrt{3}}\mqty(
		2 & 0 & 0 & 0 & -2 & 0 & 0 & 0 & -2 \\
		-1 & -i \sqrt{3} & 0 & i \sqrt{3} & 1 & 0 & 0 & 0 & -2 \\
		-1 & i \sqrt{3} & 0 & -i \sqrt{3} & 1 & 0 & 0 & 0 & -2 \\
		-1 & 1 & \sqrt{2} & 1 & -1 & \sqrt{2} & \sqrt{2} & \sqrt{2} & 0 \\
		-1 & 1 & \sqrt{2} \omega^2 & 1 & -1 & \sqrt{2} \omega^2 & \sqrt{2} \omega & \sqrt{2} \omega & 0 \\
		-1 & 1 & \sqrt{2} \omega & 1 & -1 & \sqrt{2} \omega & \sqrt{2} \omega^2 & \sqrt{2} \omega^2 & 0 \\
		-1 & -1 & \sqrt{2} & -1 & -1 & -\sqrt{2} & \sqrt{2} & -\sqrt{2} & 0 \\
		-1 & -1 & \sqrt{2} \omega & -1 & -1 & -\sqrt{2} \omega & \sqrt{2} \omega^2 & -\sqrt{2} \omega^2 & 0 \\
		-1 & -1 & \sqrt{2} \omega^2 & -1 & -1 & -\sqrt{2} \omega^2 & \sqrt{2} \omega & -\sqrt{2} \omega & 0 \\),}
	\end{equation}
	yields a point $e_p = 1,\,g_t = \frac{1}{2}$, which shows that $U'_9 P$ is actually a 2-unitary matrix related to absolutely Maximally entangled states of four systems with three levels each \cite{GALRZ15}. Indeed, among all $9!$ permutations there are $648$ permutations which yield a 2-unitary operation, meaning that not only $U'_9$ is a unitary matrix, but also certain reorderings of it yield a 2-unitary operation.
	
	\break
	
	\section{Details of the tomographic scheme}
	
	\subsection{Explicit quantum circuit for Elegant Joint Measurement} \label{app:circ}
	
	In order to derive the circuit corresponding to the EJM we will write the local matrices used in the scheme in Fig.\ref{fig:circ_gen} explicitly as

	\begin{align}
		V_1  & {=
		\mqty( e^{-\frac{i x_1}{2}-\frac{i z_1}{2}}\cos (\frac{y_1}{2}) & 
		-e^{-\frac{i x_1}{2}+\frac{i z_1}{2}} \sin (\frac{y_1}{2}) \\
		e^{\frac{i x_1}{2}-\frac{i z_1}{2}} \sin (\frac{y_1}{2}) & 
		e^{+\frac{i x_1}{2}+\frac{i z_1}{2}} \cos (\frac{y_1}{2}) ),} & 	
		V_2  & {=
		\mqty( e^{-\frac{i x_2}{2}-\frac{i z_2}{2}}\cos (\frac{y_2}{2}) & 
		-e^{-\frac{i x_2}{2}+\frac{i z_2}{2}} \sin (\frac{y_2}{2}) \\
		e^{\frac{i x_2}{2}-\frac{i z_2}{2}} \sin (\frac{y_2}{2}) & 
		e^{+\frac{i x_2}{2}+\frac{i z_2}{2}} \cos (\frac{y_2}{2}) ),} \nonumber \\
		W_1 & {=
		\mqty( e^{-\frac{i x_3}{2}-\frac{i z_3}{2}}\cos (\frac{y_3}{2}) & 
		-e^{-\frac{i x_3}{2}+\frac{i z_3}{2}} \sin (\frac{y_3}{2}) \\
		e^{\frac{i x_3}{2}-\frac{i z_3}{2}} \sin (\frac{y_3}{2}) & 
		e^{+\frac{i x_3}{2}+\frac{i z_3}{2}} \cos (\frac{y_3}{2}) ),} & 	
		W_2 & {=
		\mqty( e^{-\frac{i x_4}{2}-\frac{i z_4}{2}}\cos (\frac{y_4}{2}) & 
		-e^{-\frac{i x_4}{2}+\frac{i z_4}{2}} \sin (\frac{y_4}{2}) \\
		e^{\frac{i x_4}{2}-\frac{i z_4}{2}} \sin (\frac{y_4}{2}) & 
		e^{+\frac{i x_4}{2}+\frac{i z_4}{2}} \cos (\frac{y_4}{2}) ).}
	\end{align}
	
	Under this decomposition, basing on a non-canonical information content of the gate $U_4$,  $\qty{\pi/4,-\pi/2,-\pi/8}$ we find that the values of all the phases involved are
	\nobreak
	\begin{align*}
		\alpha & = \frac{\pi}{4}, & \beta & = 0, & \gamma & = -\frac{\pi}{2}, \\
		x_1 & = - \frac{\pi}{4}, & x_2 & = \frac{\pi}{4}, & x_3 & = 0, & x_4 & = 0, \\ 
		y_1 & = \arccos(\frac{1}{\sqrt{3}}), & y_2 & = \arccos(\frac{1}{\sqrt{3}}), & y_3 & = -\frac{\pi}{2}, & y_4 & = -\frac{\pi}{2}, \\ 
		z_1 & = \frac{\pi}{3}, & z_2 & = -\frac{\pi}{3}, & z_3 & = -\frac{\pi}{4}, & z_4 & = \frac{\pi}{4}, \\ 
	\end{align*}
	and such a transformation adds an overall phase of $\frac{\pi}{8}$ to the states, which can be neglected. The overall circuit becomes thus simplified and can be depicted as follows.
	
	\begin{figure}[H]
		\centering
		\includegraphics[width=0.75\linewidth]{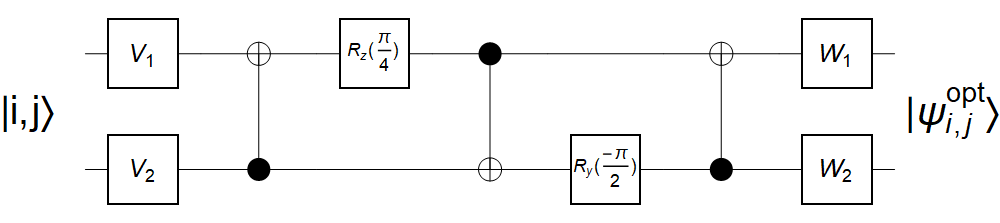}
		\caption{Circuit corresponding to the transformation between the computational basis $\ket{i,j}$ and the basis $\ket{\psi_{i,j}^{\text{opt}}}$ representing EJM, given in Eq. \eqref{eq:EJM_unitary}.}
		\label{fig:circ_opt}
	\end{figure}
	
	\subsection{Calibration data for IBM quantum computers} \label{app:calib}
	
	In Table \ref{tab:calibs} we list the calibration data for IBM quantum computers used in this work, limiting them to the relevant subsystems, that is to the first 3 qubits and CNOT gates operating between them. Based on these calibrations we estimate the success rates.
	
	\begin{table}[h!]
		\centering
		\begin{tabular}{|c|c|c|c|c|c|c|c|c||c|}
			\cline{2-10}
			\multicolumn{1}{c|}{}& \multicolumn{2}{c|}{Qubit 0} & \multicolumn{2}{c|}{Qubit 1} & \multicolumn{2}{c|}{Qubit 2} & \multirow{2}{*}{$\text{CNOT}_{01}$} & \multirow{2}{*}{$\text{CNOT}_{12}$} & \multirow{2}{*}{OSR} \\ \cline{2-7}
			\multicolumn{1}{c|}{} & R & G & R & G & R & G & & & \\ \hline
			$s$ & $10^{-2}$ & $10^{-4}$ & $10^{-2}$ & $10^{-4}$ & $10^{-2}$ & $10^{-4}$ & $10^{-3}$ & $10^{-3}$ & $10^0$ \\\hline\hline
			Melbourne & 2.05 & 5.34 & 10.2 & 15.9 & 2.02 & 9.91 & 24.11 & 14.32 & 80.66 \\ \hline
			Oursense & 2.10 & 3.19 & 2.90 & 3.59 & 1.50 & 3.12 & 6.744 & 8.335 & 92.18\\ \hline
		\end{tabular}
		\caption{Calibrations of the relevant procedures of quantum computers used, as found on the days of measurements, collecting error rates readout (R), one-qubit gates (G) and CNOT gates. Errors for CNOT gates operating on two qubits given by the subscript are found to be symmetric with respect to exchange of the two qubits. For clarity all the errors are scaled by factor $s$ given in the first row. Overall success rate (OSR) was estimated basing on the quantum circuit corresponding to EJM and the standard Bell pair preparation subroutine.}
		\label{tab:calibs}
	\end{table}
	
	\subsection{Results of the tomography procedure} \label{app:results}
	
	Results of measurements listed in Table \ref{tab:meas} have been collected from a total of 18 runs of two different IBM quantum computers. Simulations give the average mean deviation $\sqrt{\Tr(\Delta\sigma_{\rm sim}^2)}\approx 0.027$. The real-life experimental data give the mean deviation greater by almost one order of magnitude, $\sqrt{\Tr(\Delta\sigma_{\rm exp}^2)}\approx 0.177$.
	
	\begin{table}[h!]
		\centering
		\footnotesize
		\begin{tabular}{|c|c|c|c|c|c||c|c|c|c|c|}
			\cline{2-11}
			\multicolumn{1}{c|}{}& $\ket{00}$ & $\ket{01}$ & $\ket{10}$ & $\ket{11}$ & $\sqrt{\Tr(\Delta\sigma^2)}$ & $\ket{00}$ & $\ket{01}$ & $\ket{10}$ & $\ket{11}$ & $\sqrt{\Tr(\Delta\sigma^2)}$ \\ \cline{2-11}
			\multicolumn{1}{c|}{}& \multicolumn{5}{c||}{State $\ket{0}$}& \multicolumn{5}{c|}{State $\ket{1}$} \\ \hline
			Predicted & 46.7 & 17.8 & 17.8 & 17.8 & - & 3.3 & 32.2 & 32.2 & 32.2 & - \\ \hline
			Bell simulated & 46.6 & 17.9 & 17.8 & 17.7 & 0.005 & 3.4 & 31.6 & 33.7 & 31.3 & 0.051 \\
			Bell Melbourne & 46.6 & 19.9 & 18.5 & 15.1 & 0.099 & 10.3 & 30.7 & 31.1 & 27.9 & 0.239 \\
			Bell Oursense & 45.4 & 16.5 & 19.4 & 18.7 & 0.073 & 6.2 & 25.0 & 33.2 & 35.6 & 0.240 \\ \hline
			\multicolumn{9}{c}{} \\ \cline{2-11}
			\multicolumn{1}{c|}{}& $\ket{00}$ & $\ket{01}$ & $\ket{10}$ & $\ket{11}$ & $\sqrt{\Tr(\Delta\sigma^2)}$ & $\ket{00}$ & $\ket{01}$ & $\ket{10}$ & $\ket{11}$ & $\sqrt{\Tr(\Delta\sigma^2)}$ \\ \cline{2-11}
			\multicolumn{1}{c|}{}& \multicolumn{5}{c||}{State $\ket{+}$}& \multicolumn{5}{c|}{State $\ket{-}$} \\ \hline
			Predicted & 25.0 & 45.4, & 14.8 & 14.8 & - & 25.0 & 4.6 & 35.2 & 35.2 & - \\ \hline
			Bell simulated & 25.6 & 44.3 & 14.8 & 15.3 & 0.038 & 24.9 & 4.7 & 34.8 & 35.6 & 0.015 \\
			Bell Melbourne & 31.6 & 41.8 & 13.0 & 13.6 & 0.222 & 26.2 & 9.0 & 34.9 & 29.9 & 0.199 \\
			Bell Oursense & 28.7 & 37.3 & 16.1 & 17.9 & 0.270 & 21.9 & 6.5 & 34.9 & 36.6 & 0.110 \\ \hline
			\multicolumn{9}{c}{} \\ \cline{2-11}
			\multicolumn{1}{c|}{}& $\ket{00}$ & $\ket{01}$ & $\ket{10}$ & $\ket{11}$ & $\sqrt{\Tr(\Delta\sigma^2)}$ & $\ket{00}$ & $\ket{01}$ & $\ket{10}$ & $\ket{11}$ & $\sqrt{\Tr(\Delta\sigma^2)}$ \\ \cline{2-11}
			\multicolumn{1}{c|}{}& \multicolumn{5}{c||}{State $\ket{\otimes}$}& \multicolumn{5}{c|}{State $\ket{\odot}$} \\ \hline
			Predicted & 25.0 & 25.0 & 7.3 & 42.7 & - & 25.0 & 25.0 & 42.7 & 7.3 & - \\ \hline
			Bell simulated & 25.2 & 25.1 & 7.6 & 42.0 & 0.022 & 25.4 & 25.3 & 41.8 & 7.4 & 0.028	\\
			Bell Melbourne & 27.3 & 25.6 & 12.8 & 34.2 & 0.293 & 29.1 & 24.9 & 37.6 & 8.5 & 0.187 \\
			Bell Oursense & 24.5 & 22.0 & 8.3 & 45.1 & 0.114 & 26.5 & 22.8 & 43.5 & 7.2 & 0.078 \\ \hline
		\end{tabular}
		\caption{Percentage of states observed in each experiment, together with the average Hilbert-Schmidt distance of the reconstructed state $\sigma_\text{reco}$ from the accurate state $\sigma_\text{th}$.}
		\label{tab:meas}
	\end{table}	
	
	\normalem
	\bibliographystyle{ieeetrantmp}
	\bibliography{biblio}
	
\end{document}